\documentclass{optica-article}

\journal{opticajournal} 

\articletype{Research Article}

\begin{document}

\title{First Implementation and Characterization of a Fiber Amplifier in a Gravitational Wave Detector}

\author{Margherita Turconi \authormark{1}, Marina Trad Nery \authormark{1}, Remi Soulard \authormark{1,3}, Matthieu Gosselin \authormark{2} , Piernicola Spinicelli \authormark{2}, Suzanne Assis de Souza Melo \authormark{2}, Camilla De Rossi \authormark{2}, Antonino Chiummo \authormark{2}, Mourad Merzougui \authormark{1}, Jean-Pierre  Coulon \authormark{1}, Nelson Christensen \authormark{1} and Walid Chaibi \authormark{1}} 

\address{\authormark{1}Université Côte d’Azur, Observatoire de la Côte d’Azur, CNRS, Artemis, 06304 Nice, France\\
\authormark{2}European Gravitational Observatory (EGO), Via E Amaldi, 5, 56021 Cascina (PI), Italy \\
\authormark{3} Current address: Centre de Recherche sur les Ions, les Matériaux et la Photonique (Cimap), UMR 6252, 5 Boulevard Maréchal Juin, 14050 Caen, France.}

\email{\authormark{*}marina.trad-nery@oca.eu} 


\begin{abstract*} 

Gravitational-wave detectors rely on large-scale interferometers operating at exquisite sensitivities, where the laser source plays a central and critical role. The sensitivity of these instruments is directly influenced by the power level and noise properties of the injected light.
Fiber amplifiers are a promising technology that are currently considered for the next generation of detectors because of their potential in laser power scaling and their superior output beam quality.
In the context of the Virgo gravitational wave detector, we report the first implementation of a fiber amplifier into Virgo's injection system as a high-power laser source. We demonstrate a 130 W output beam that fulfills the requirements in terms of frequency noise, power noise, and beam-pointing noise. Additionally, we also show the performance of the stabilization loops that enable the commissioning of the Virgo detector. With more than 17,000 hours of operation, this work represents the longest continuous characterization and operation of a high-power and single-frequency fiber amplifier developed for gravitational-wave detection. 

\end{abstract*}

\section{Introduction}

Since the first direct detection of gravitational waves in 2015, the LIGO–Virgo-KAGRA Collaboration has detected hundreds of gravitational-wave signals. To date, 390 detections have been announced, including binary black hole systems spanning a wide range of masses and mass ratios, as well as black hole–neutron star and binary neutron star mergers \cite{GWTC1,GWTC2,GWTC3,LIGOScientific:2025slb,LIGOScientific:2026sit}. This diversity, combined with the observation of electromagnetic counterparts across a broad range of wavelengths, has opened a new era in astrophysics, known as multi-messenger astrophysics \cite{multimess}. These developments have driven significant efforts toward the development of third-generation gravitational wave (GW) detectors, which have a substantial technological leap compared to the current generation. A key strategy to improve their sensitivity is to increase the laser power. For example, the European project Einstein Telescope \cite{ET2020} foresees increasing the power at the interferometer input from the 60-75 W used in LIGO in O4 \cite{LIGO_O4} to 500 W at a wavelength of 1064 nm. Such an increase is extremely challenging given that the high power laser beam must also be highly stable in the short and long term, linearly polarized, single frequency, and single mode. \\    

So far GW detectors have traditionally relied on solid state laser technology implemented either on injection-locked topologies \cite{Kwee2012} or in a Master Oscillator Power Amplifier (MOPA) configuration \cite{neovan,LIGO_O3_laser,LIGO_O4_laser}. However, solid state amplifiers suffers from thermally induced optical deformation, which degrades beam quality at power levels envisioned by future detectors. For example, up to 195 W were demonstrated with two sequential amplifiers, with 11\% higher order mode content \cite{LIGO_O4_laser}, while further power scaling leads to reduced spatial purity and complicates the management of unused power. A promising alternative are fiber amplifiers used in MOPA configuration thanks to their compactness, high overall amplification efficiency and beam quality. Narrow linewidth linearly polarized laser using ytterbium-doped fibers have been of increasing interest and have reached a maximum power of 7 kW~\cite{Pengfei2022}. For GW detectors, fiber amplifiers are still limited to hundreds of Watts since an ultra low noise single frequency beam is needed, and high power amplifiers are limited by two specific non linear effects: (i) the stimulated Brillouin scattering (SBS)\cite{SBS} which limits the output power and increases the power noise in the MHz frequency range \cite{SBS_noise,Guiraud2016}, and the (ii) thermally induced mode instability (TMI) which is responsible for the degradation of the output mode and a substantial increase of the noise~\cite{Wirth2010,Zhao2018}. Considerable R\&D has demonstrated promising performance ~\cite{CBC_Artemis,Wellmann2019,Wellmann2020,Dixneuf2020}, and no insurmountable technical barriers to integration into GW detectors have been encountered. The best results in terms of stable linearly-polarized SBS and TMI free emission and robustness have been obtained with polarization maintaining large mode area (LMA) ytterbium doped fibers \cite{Wellmann2019,Wellmann2020,Dixneuf2020}. Another alternative to reach higher powers while preserving beam quality, is the technique of coherent beam combination, which has been proposed and demonstrated, although not yet implemented in GW detectors \cite{CBC_Artemis,CBC400,Bode2023}. For Einstein Telescope, a combination of fiber amplifiers and coherent beam combination is currently foreseen to achieve the required power while preserving beam quality.

Within the framework of the Advanced Virgo+ project~\cite{Acernese2023}, a decision was made to integrate a high-power fiber amplifier to increase the power at the output of the laser system to 130 W. This decision was taken due to the already good results obtained with this technology and also due to their compactness, simple operation, and long term reliability. In this paper, we present the validation tests conducted on this amplifier and report on the results of its implementation within the Virgo detector. We start by describing the laser system at Virgo, then we show the validation tests of the fiber amplifier conducted at the Artemis laboratory, and finally the results on the installation and comissioning of the amplifier at the Virgo detector.

\section{The Virgo Pre-Stabilized Laser subsystem}
\label{sec:PSL}
The Pre-Stabilized Laser (PSL) is a subsystem of the Injection system (ISYS) which provides the interrogation beam to the interferometer. The role of the ISYS system is to insure a high power laser beam with:
\begin{itemize}
    \item a first stage of frequency stabilization, and the actuation for a second stage of frequency stabilization on the interferometer common mode at a level of a few $10^{-8}\,\textrm{Hz/Hz}^{1/2}$ on the detection frequency band (10 Hz-10 kHz),
    \item a relative power (power noise over mean power) stabilization at the level of a few  $10^{-9}\,\textrm{Hz}^{-1/2}$ at frequencies around 10 Hz \cite{ClevaPstab},
    \item a transverse single fundamental mode with a very low beam pointing noise,
    \item and different phase modulation sidebands to be used for stabilization purposes required by the detector.
\end{itemize}

More details and a schematics on the ISYS system will be given at Sec.\ref{virgo_injection_system}. For now, we focus on the PSL subsystem of ISYS, which provides the high-power laser source, power stabilization, and frequency actuation while meeting stringent requirements on frequency noise, power noise, beam-pointing noise, and long-term stability. The high power beam is produced through the amplification of a master laser/seeder which is a Non-Planar-Ring-Oscillator (NPRO) at 1064 nm from COHERENT\textregistered{}. The NPRO has been used in GW detectors for over 30 years, since it delivers up to a few watts of exceptionally low noise beam both in terms of frequency and power \cite{Vahlbruchnpro}. It is also highly reliable and can operate for years without undergoing any failure.
Building on this robust low-noise laser source, a major challenge for PSL systems over the past two decades has been to increase the optical power while preserving single-frequency, single-transverse-mode operation and maintaining compatibility with the various stabilization control loops. 

For the first Virgo project, the PSL was required to provide a 10 W laser beam which was generated by an injection-locked cavity called slave laser (SL) \cite{ISYS}. The same beam was used during the following detector upgrades, namely Virgo+ and Advanced Virgo (AdV), but as a seeder for a solid state MOPA that provided 60 W and 100 W, respectively. The next project upgrade, called AdV+, is subdivided into two phases : (i) Phase 1 where the PSL output power was required to be around 80 W in order to provide 40 W at the interferometer input (due to the optical losses through the PSL and the injection system), and (ii) Phase 2 for which an upgrade of the system was planned to increase the PSL output power to 140 W, ensuring that the power available for the interferometer exceeds 80 W. For this a fiber technology based MOPA was adopted. The fiber amplifier was manufactured by the company TOPTICA\footnote{https://www.toptica.com/} (formerly Azurlight Systems) and its output power could reach up to 130 W. Before installing a new subsystem on the detector, a protocol of validation tests have to be followed to ensure its compatibility with the detector's requirements and also with its infrastructure. In the next section, we present the results of those tests which were performed in the Artemis laboratory. 

\section{Validation tests of the MOPA high power fiber amplifier}

\begin{figure}[ht!]
\centering
    \includegraphics[width=\textwidth]{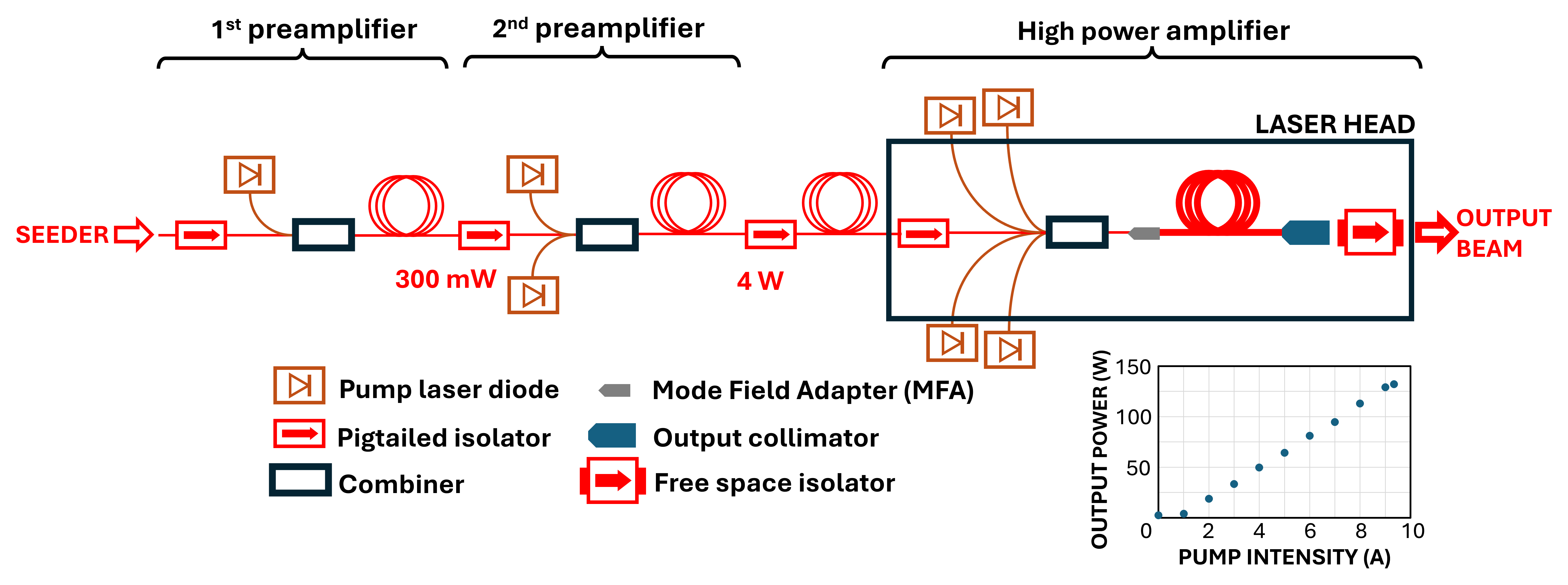}
\caption{Simplified architecture of the fiber amplifier. The system consists of three cascaded amplification stages, with the final high-power stage integrated into the laser head, whose output beam passes through a free-space optical isolator. The insert in the bottom right corner shows the output power as function of the last stage pump current. Details in the text.}
\label{figure_amplifier_architecture}
\end{figure}

We start by describing the fiber amplifier architecture which, according to the information provided by the manufacturer, is displayed in Fig.\ref{figure_amplifier_architecture}. 
The injected seeder, an NPRO laser with a power between 10-100 mW, is amplified in three stages: the first preamplifer provides 300 mW which are amplified up to 4 W by a second preamplifer. The fiber mode is then adapted to the final amplifying fiber (high power amplifier stage) with a Mode Field Adapter (MFA). The high power output beam is collimated and then passes through the free space output isolator. Among other reasons, pigtailed isolators are interspersed between each amplifying stage to avoid SBS. Due to constrains related to the installation and the use of the fiber amplifier in the Virgo Laser Lab, the second preamplifier output fiber was lengthened up to 2 m and an additional pigtailed isolator was placed. 

The fiber amplifier full output power was measured with a calorimeter while increasing the pump's current intensity of the final stage (see insert in Fig.\ref{figure_amplifier_architecture}). At 0 A, the measured output power was 2 W. From 1 A to 9 A, the output power increased linearly with a slope of $\approx$ 16 W/A, right before reaching a saturation behavior. The maximum power reached was 131 W at the maximum current, which was limited to 9.5 A.

\subsection{Beam modal analysis}
The PSL is required to provide a beam with a spatial quality as close as possible to a pure fundamental Gaussian mode, preferably higher than 90\%. The main reason is to reduce the reflected laser power by a pre-mode cleaner cavity, which must be safely dumped to minimize scattered light and thermal load on the laser bench.

\subsubsection{Preliminary beam analysis}
\begin{figure}[ht!]
\centering\includegraphics[trim= 0cm 3cm 0cm 3cm,clip,width=11cm]{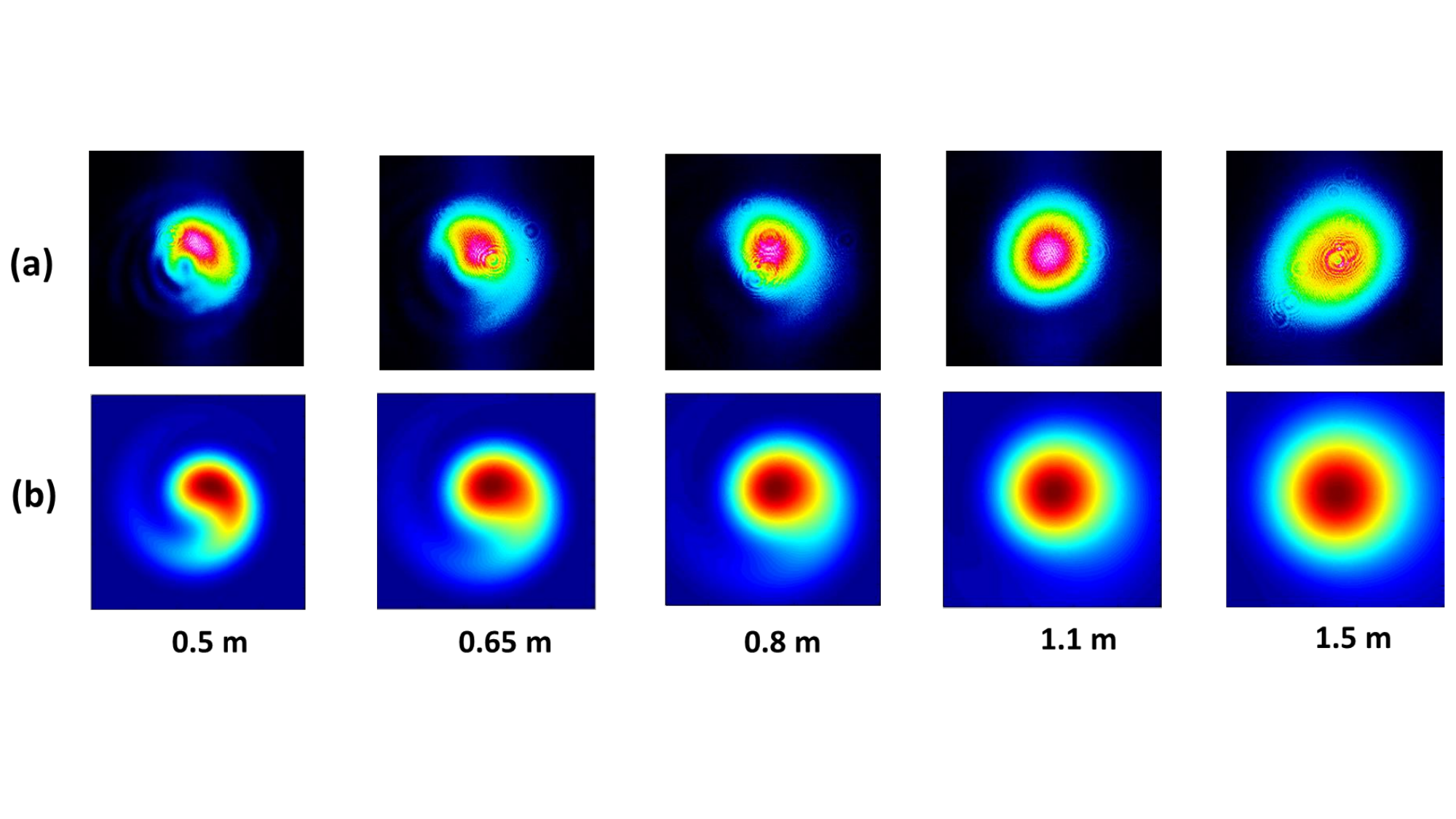}
\caption{(a) Behavior of the output beam shape while propagating out of the laser head; $z=0\,\textrm{m}$ corresponds to the position of the laser head output window. (b) Simulation model based on an additional donuts mode propagating within the cladding, which qualitatively reproduce the output mode behavior. The size and waists of both fundamental and donuts modes, as well as the collimator characteristics were set as parameters that were adjusted to reproduce the output beam behavior.}
\label{figure_modal_simulation}
\end{figure}

As a first characterisation of the modal behavior of the fiber amplifier output beam, its transversal mode was imaged on a WinCamD camera from DataRay  Inc.\textregistered{}, at different positions from the output window of laser head. The obtained images are shown in row (a) of Fig.\ref{figure_modal_simulation}. They show that the beam shape is not purely Gaussian at the output of the laser head but becomes more Gaussian-looking while propagating. A possible phenomenological explanation was simulated and is shown in row (b). It illustrates the propagation of a beam profile which has 90\% of its power into the fundamental mode and 10\% into a donut mode formed by the superposition of $\textrm{HG}_{10}$ and $\textrm{HG}_{01}$ with a $\pi/2$ phase shift. The donut mode is assumed to have a slightly larger beam radius than the fundamental core mode, modeling light that partially leaks into the cladding during propagation. Near the waist, the donut mode distorts by interference the central part of the fundamental mode. Whereas in the far field, it diverges more rapidly than the fundamental mode and its contribution to the central part vanishes resulting in a mode which is mainly fundamental. This simulation reproduces qualitatively the evolution of the measured output mode and shows how the mode can be distorted due to power in higher order modes.

\begin{figure}[ht!]
\centering\includegraphics[trim= 0cm 1cm 0cm 1cm,clip,width=12cm]{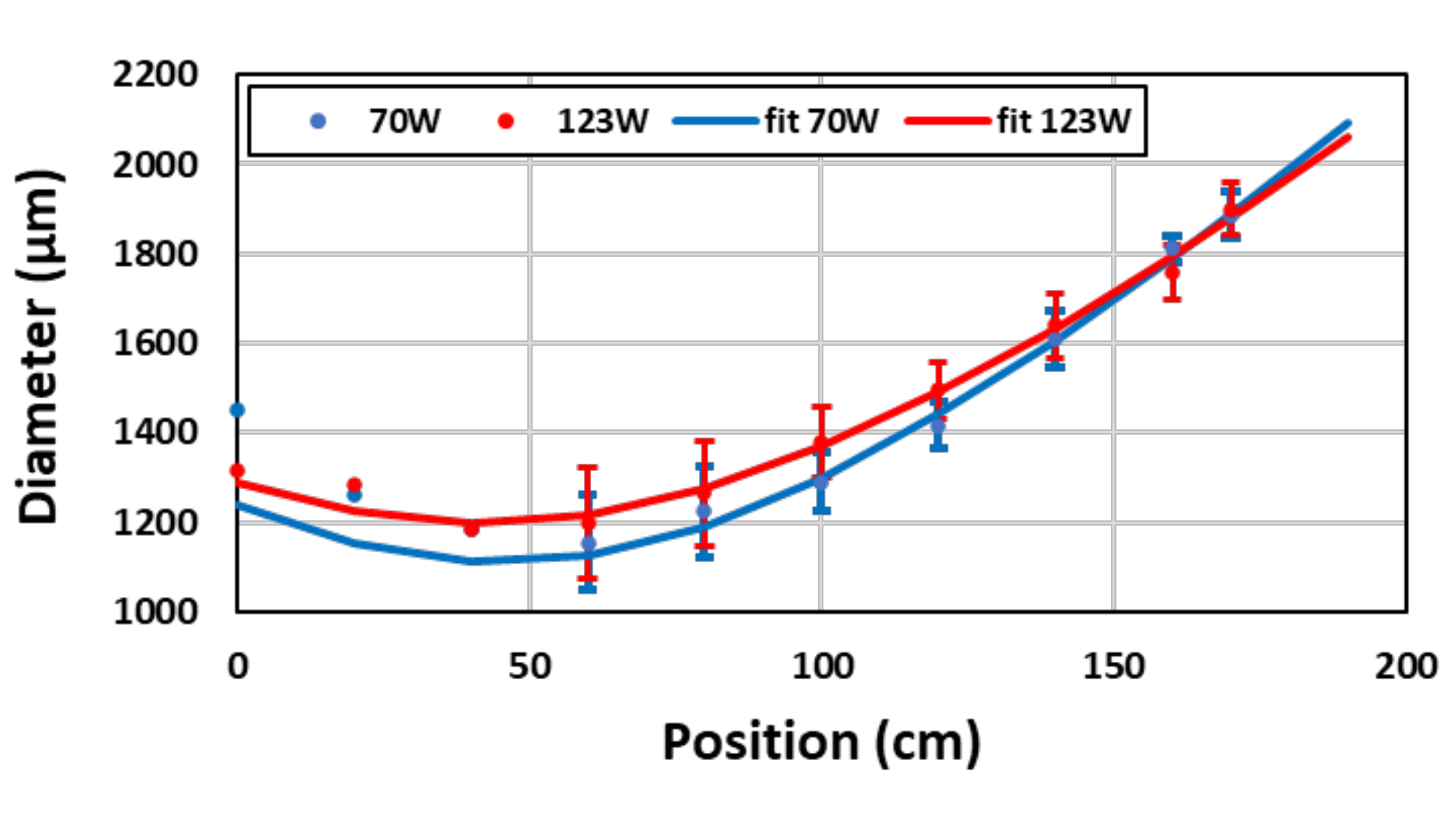}
\caption{Measurement (dots) and fit (solid lines) of the beam diameter evolution with position from the fiber amplifier laser head.}
\label{figure_modal_fit}
\end{figure}

The evolution of the propagating mode along the distance to the laser head is shown in Fig.~\ref{figure_modal_fit} with error bars corresponding to a 95\% confidence interval, which was estimated with the WinCam D camera software. The profile allows to estimate the waist average radius $w_0$ and its position $z_0$, necessary to design a telescope for the beam injection into a reference cavity (described in the next paragraphs). For each position, the image was 2D fitted as an elliptical Gaussian mode. The average radius was determined from the minor and major radii. Since the beam was not purely Gaussian, the contribution of higher order modes caused an error during the estimation, which was taken into account during the $\chi^2$ fit procedure. This procedure was conducted for 70 W and 123 W and in both cases, the radius measurements were not considered for $z<60$ cm as the beam was mainly distorted. The fit procedure outcome was:
\begin{itemize}
    \item for 70 W output power $z_0=45\pm 2.2\,\textrm{cm}$ and $w_0=555\pm 20\,\mu\textrm{m}$, 
    \item for 123 W output power $z_0=42\pm 2.8\,\textrm{cm}$ and $w_0=600\pm 24\,\mu\textrm{m}$.
\end{itemize}
These results show an overlap between the beams' characteristics for both powers and the matching telescope to the reference cavity was not required to be adjusted when the output power was changed. That is a practical advantage compared to solid state amplifiers in which the output mode changes with the output power.

\subsubsection{Coupling efficiency into a reference cavity}
\begin{figure}[ht!]
\centering\includegraphics[width=12cm]{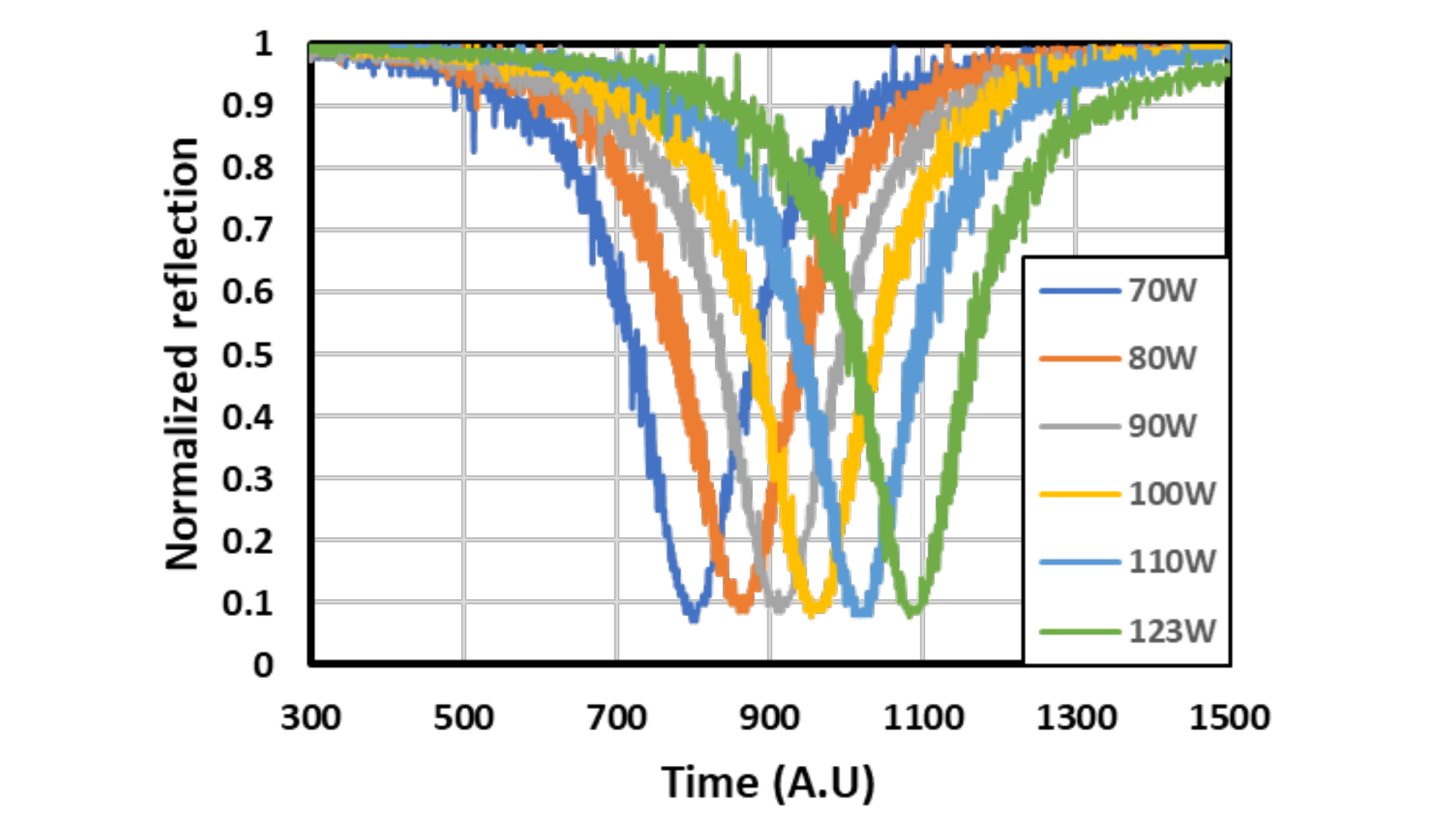}
\caption{Reflection signals of the reference cavity during laser frequency scans for different powers. The frequency scans were performed at the same rate for all measurements and signals were normalized and shifted for better visualisation. The photodiode offset was subtracted from all signals.}
\label{figure_modal_reflection}
\end{figure}
The high power beam was injected into an ultra stable low thermal expansion and non degenerate triangular reference cavity which is a twin cavity of a reference cavity installed in Virgo. The cavity spacer is made of ceramics (Zerodur) spacer, and has corner flat mirrors seen with an incidence of 44°30'. Its free spectral range (FSR) is 470 MHz which corresponds to a cavity length of 32 cm. As a triangular cavity, it has two different Finesses depending on the impinging polarization: $\simeq 950$ for the $p$ polarization and $\simeq 32000$ for the $s$ polarization, measured at the Virgo site \cite{Canuel2013}. The coupling efficiency was estimated from the ratio of the reflected power off and on resonance while scanning the seed laser frequency, for amplifier output powers between 70 W and 123 W.
As shown in Fig.~\ref{figure_modal_reflection}, the coupling efficiency was 91.5\% and was power independent within 1\%. This result is in agreement with the higher order mode content used in the simulation of Fig. \ref{figure_modal_simulation}.

\subsubsection{Modal content}
\begin{figure}[ht!]
\centering\includegraphics[width=11cm]{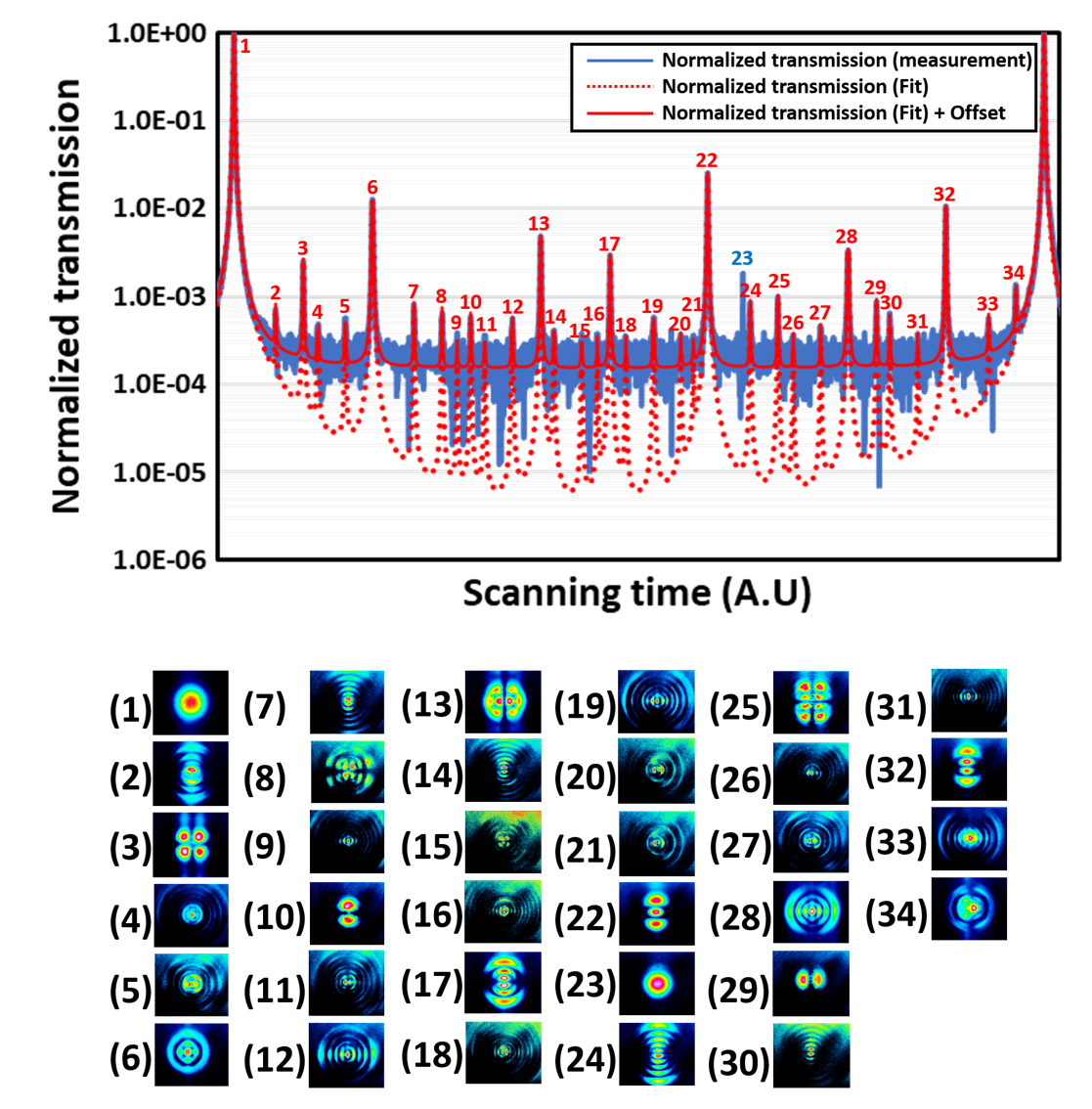}
\caption{Modal content of the high power amplifier beam measured in transmission of the reference cavity while scanning the laser frequency. All resonances were normalized with respect to the fundamental mode. An offset of $1.5\times 10^{-4}$ was taken into account in red curve, which corresponds to the electronic offset of the photodiode. Several artificial peaks due to concatenation were ignored in the fits. Mode shapes are given on the right panel and were measured by a camera on the transmitted beam.}
\label{figure_modal_content}
\end{figure}
The ultimate analysis of the beam quality was based on the measurement of the beam modal content, which was performed by scanning the seed laser frequency and measuring the higher order mode power transmitted by the triangular cavity at its resonance. In this case, observing the cavity transmission allows the identification of each mode constituting the beam ~\cite{beam_quality_jitter_AEI}. Due to the limited dynamic of the NPRO PZT, the crystal temperature was adjusted to cover the whole cavity FSR. The set of modes' resonances is shown in Fig.~\ref{figure_modal_content} and is the concatenation of 20 different spectra, each one obtained for a different laser crystal temperature. Here below few important remarks:
\begin{itemize}
    \item A fit procedure was conducted to identify each mode contribution. Due to the tilted incidence on the curved mirror of the reference cavity~\cite{triangular_cavity}, all modes were astigmatic which lifts the degeneracy of the Hermite Gauss modes of the same $m+n$ (with the notations of~\cite{Siegman1986}). Several peaks are doubled and could not be resolved with the camera but were taken into account by the fit procedure. The fundamental mode contribution was then inferred to be 92.6\% with an accuracy of the order of $10^{-4}$. This in agreement with the measurement based on the reflected beam if a round trip loss of 200 ppm is assumed, which is an acceptable value for this cavity.
    \item The misalignment is given by HG$_{0,1}$ for the cavity $y$ axis (peak (10) in Fig.~\ref{figure_modal_content} ) and HG$_{1,0}$ for the cavity $x$ axis (peak (29)) and was limited to 0.13\%.
    \item The peak (23) corresponds to the fundamental mode in the $s$ polarisation, which resonates at high finesse in the triangular cavity. The beam polarization was optimized to $p$; the residue of $s$ polarization is due to the defect of the used polarization optics and was limited to 0.2\%.
   \item The peak (22) at 2.7\% represents the HG$_{0,2}$ mode, which was the vertical contribution to the mismatching Laguerre Gauss mode LG$_{1,0}$. The horizontal contribution HG$_{2,0}$ was not measurable showing that the injected beam was astigmatic/elliptical and that the matching was optimized for the horizontal axis. 
\end{itemize}

\subsection{Frequency noise}
\label{Section_frequency_noise}
\begin{figure}[ht!]
\centering\includegraphics[width=0.9\textwidth]{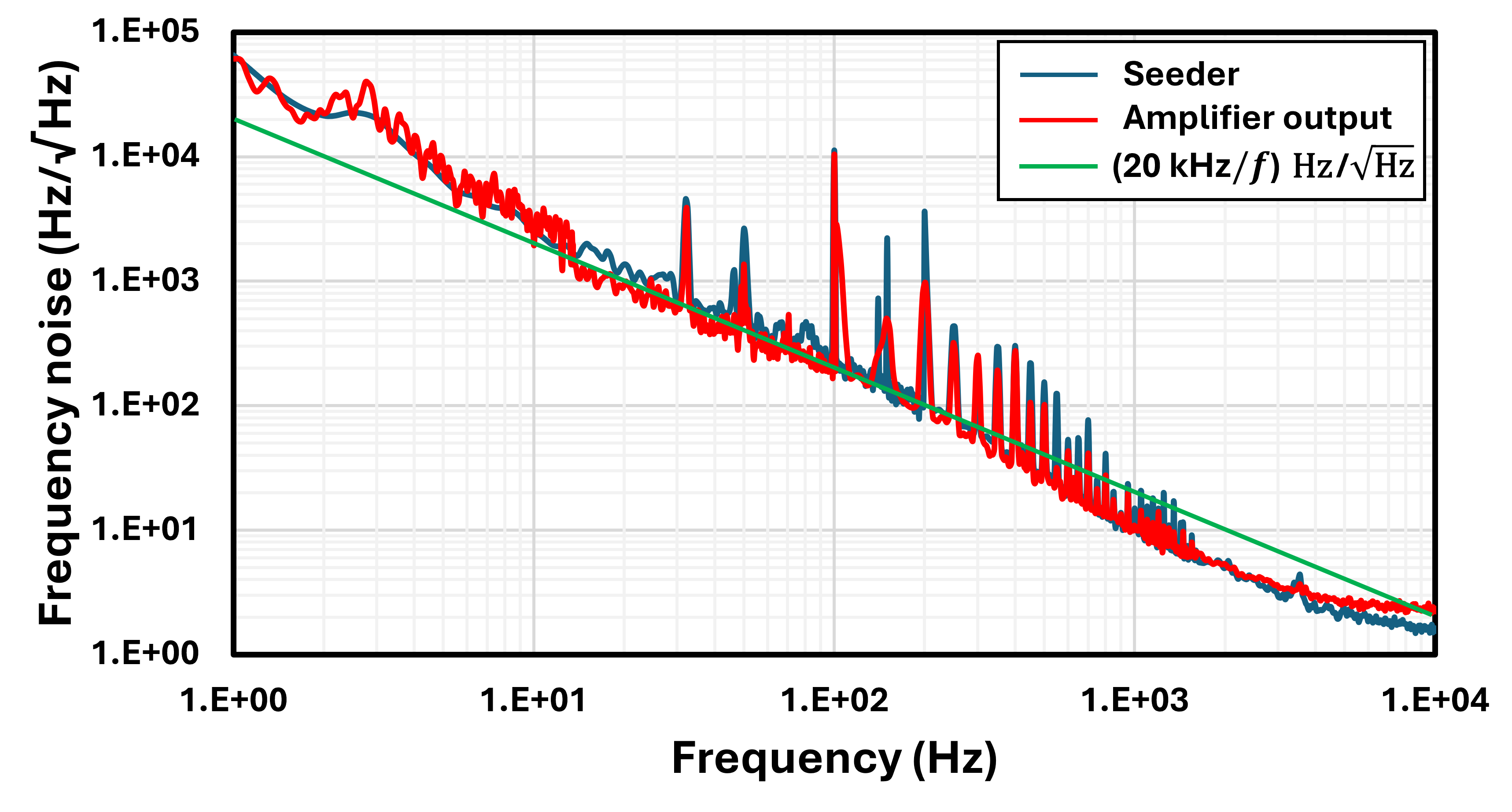}
\caption{ASD of the amplifier frequency noise (red) versus NPRO seeder frequency noise (blue), and typical NPRO frequency noise $1/f$ behavior (green). }

\label{figure_frequency_noise}
\end{figure}
 The measurement of the fiber amplifier frequency noise was conducted using the same reference cavity used for the modal content analysis, as a frequency reference. The frequency noise measurement was performed by stabilizing the laser frequency to the reference cavity using a standard Pound-Drever-Hall (PDH) locking technique \cite{Drever1983} with a Unity Gain Frequency (UGF) of 15 kHz, and measuring the correction signal. The locking used the internal laser PZT as a frequency actuator which was fast enough to reach the required locking bandwidth. No actuation on the laser crystal temperature was required since a few seconds were enough to acquire the correction signal. Below the UGF, the amplitude spectral density (ASD) of the frequency noise $\delta \nu$ of the laser with respect to the reference cavity is:
\begin{equation}
\delta \nu \simeq \gamma \times \delta C ,
\end{equation}where $\gamma$ is the frequency response of the PZT actuator, measured to be 2 MHz/V in agreement with the laser supplier's specifications, and $\delta C$ is the ASD of the correction signal in V$/\sqrt{\textrm{Hz}}$.
First, the frequency noise of only the NPRO (seed) laser was measured (blue curve in Fig.\ref{figure_frequency_noise}), which is in agreement with the expected behavior of $\delta\nu \simeq \left(20\,\textrm{kHz}/f\right)\textrm{Hz}/\sqrt{\textrm{Hz}}$ \cite{DAy1991,Millo2014} (green curve in Fig.~\ref{figure_frequency_noise}). Then a sampled beam of the amplifier output was injected into the reference cavity and the frequency noise was re-measured (red curve in Fig.~\ref{figure_frequency_noise}), showing that no additional frequency noise was added by the fiber amplifier in the Virgo detection band from 1 Hz to 10 kHz, within the measurement sensitivity. A slight discrepancy is observed at high frequencies $f>5$ kHz that could be explained by a difference between the lock bandwidths between the two measurements that was not carefully controlled.

\subsection{Relative Power Noise}

\begin{figure}[ht!]
\centering\includegraphics[width=\textwidth]{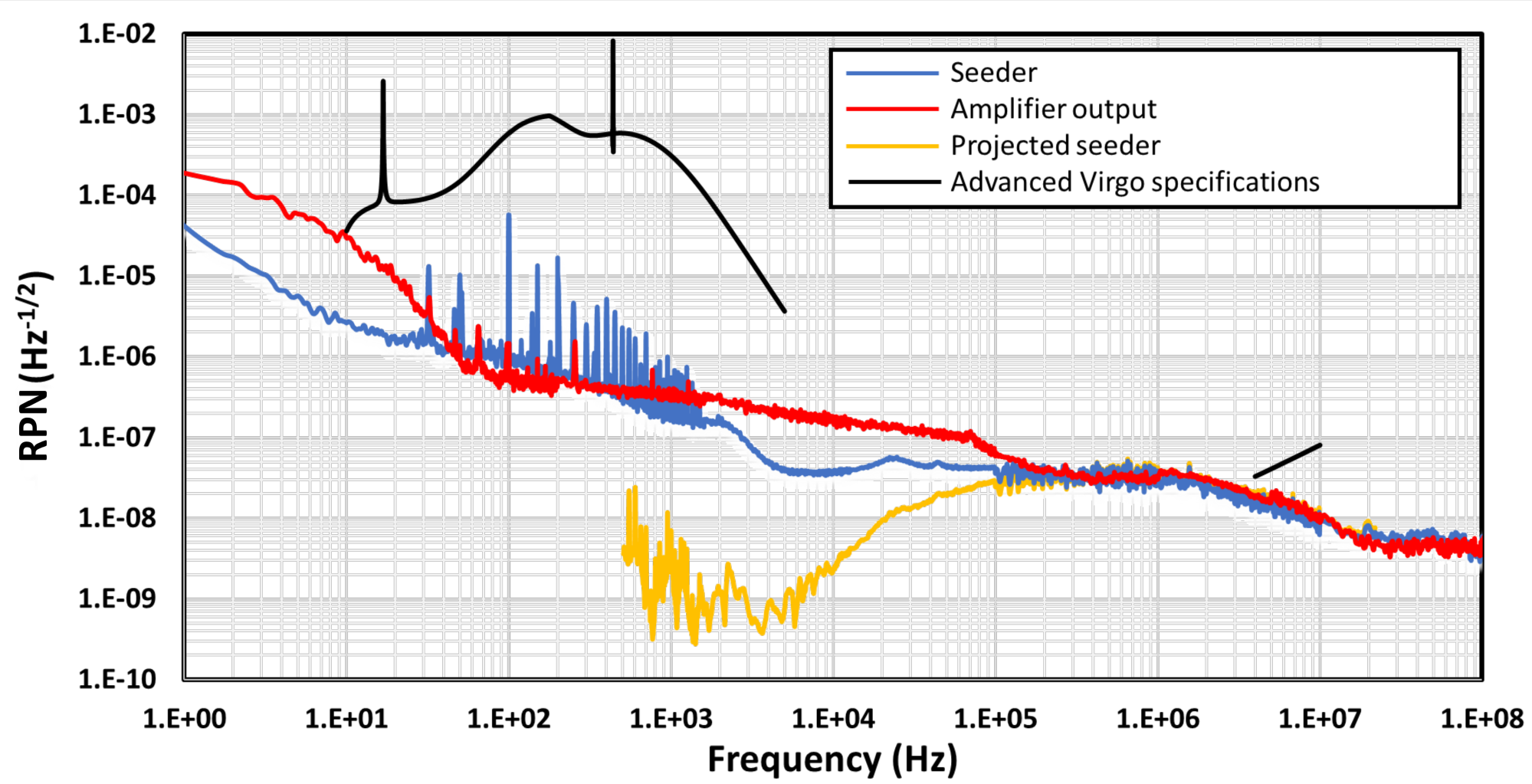}
\caption{ASD of the relative power noise of the seeder (blue curve) and the amplifier output (red). The figure also shows the Adv. Virgo requirements for the relative power noise (RPN) after the fiber amplifier (black curve) and a transfer function from RPN seeder to RPN fiber amplifier based projection (yellow curve). 
}
\label{figure_RIN}
\end{figure}
Power fluctuations at the interferometer input couple to the interferometer sensitivity via different paths. A direct coupling path in AdV+ is via the DC-read out technique in which the interferometer is slighlty detuned from the dark-fringe, resulting in a small power beam at the output \cite{DCreadout}. High frequency power noise can also couple to the readout via coupling effects involving phase modulation sidebands.

\begin{figure}[ht!]
\centering\includegraphics[trim= {1cm 2cm 1cm 0.5cm}, clip, width=\textwidth]{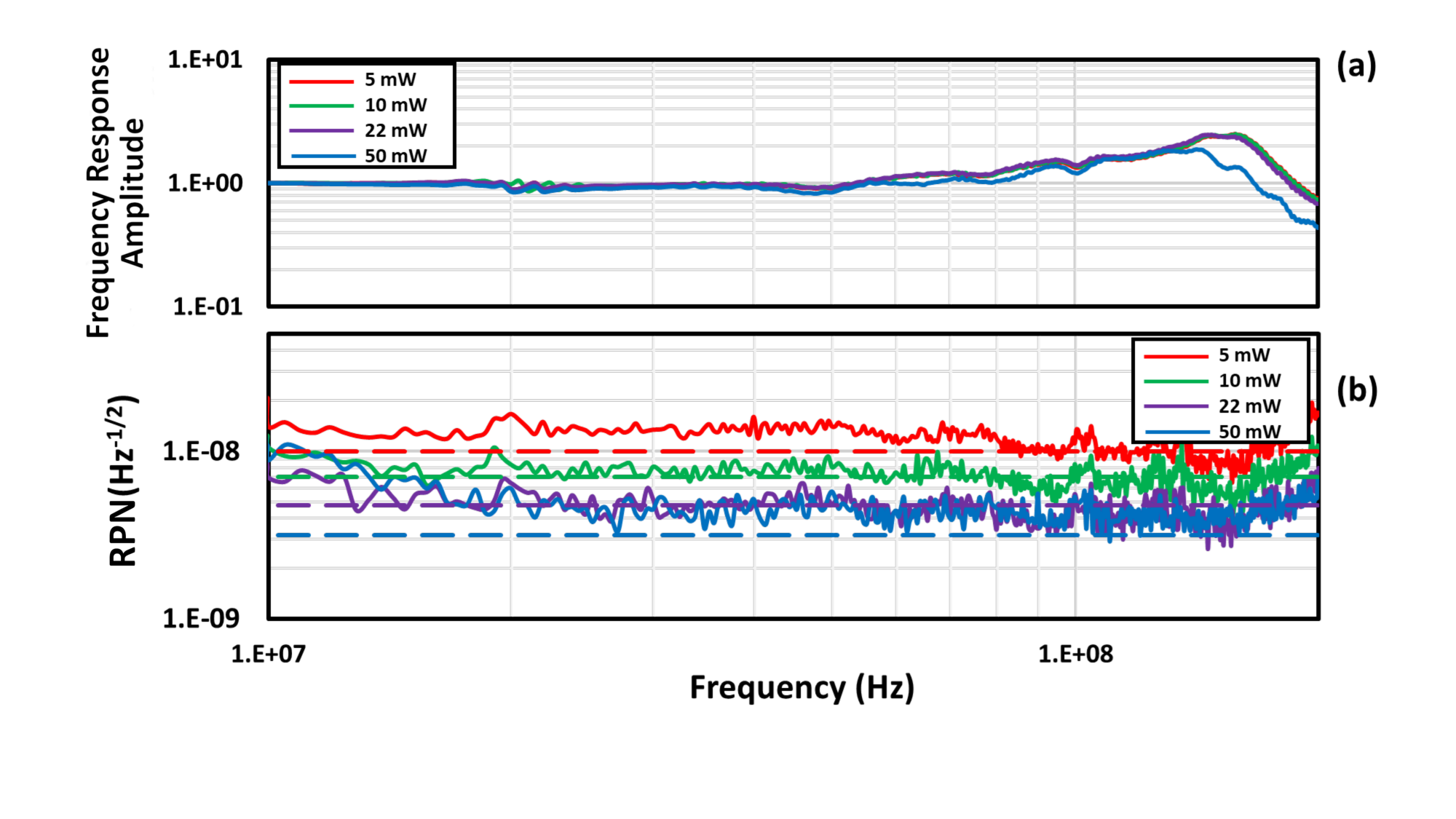}
\caption{(a) Normalized amplitude of the high frequency saturated photodiode's frequency response with respect to the impinging power. The 500 $\mu$m diameter photodiode had a saturation power of 5 mW. Up to 22 mW, the response remains unchanged with a maximum at 150 MHz. For 50 mW power, the response is reduced between 130 and 200 MHz. (b) RPN of the seeder measured with the saturated photodiode for different impinging powers (straight lines) compared to the corresponding relative shot noise levels (dashed lines). From 5 to 22 mW, the RPN drops with increasing power showing that the measurement was limited by shot noise. The RPN at 50 mW is quantitatively the same as 22 mW, and the measurement is no longer limited by the relative shot noise. The RPN increases for frequencies higher than 150 MHz for all impinging powers, which are likely to be explained by a discrepancy during the measurement of the photodiode frequency response. Hence, only measurements for $f<100$ MHz were considered.}
\label{figure_RIN_HF}
\end{figure}

The relative power noise (RPN) requirements for the beam at the output of the fiber amplifier are shown as the black curve in Fig.~\ref{figure_RIN}. They were calculated by combining the target interferometer sensitivity with a model of the dominant power noise coupling mechanisms, accounting for the low-pass filter responses of the pre-mode cleaner and input mode cleaner cavities as well as the suppression provided by the power stabilization loop (these components will be described in details in Sec.\ref{virgo_injection_system}) \cite{ClevaPstab}.  
Measurements of the RPN of the seeder and of the amplifier output beam are shown in Fig.~\ref{figure_RIN}. These measurements required two different procedures, depending on the frequency range of interest:
\begin{itemize}
    \item At low frequencies, the power noise of a pick-off beam was measured with the following precautions to reduce air turbulence coupling below 100 Hz: the clean air flow was stopped during the measurement, the pick-off laser beam was covered from the laser head down to the photodiode such that it was shielded against the air turbulence, and the main beam was sent outside the air flow cabinet to remove turbulence generated when the high power beam was dumped.
    \item At frequencies above 1 MHz, the RPN can be limited by the relative shot noise of the pick-off beam if its power is not high enough. To verify that the measurement was not shot-noise limited, the pick-off beam power $P$ was gradually increased until the RPN no longer followed the $1/\sqrt{P}$ relative shot-noise scaling. 
	This procedure applied to the seeder is shown in Fig.~\ref{figure_RIN_HF}b where one can see the RPN converging towards a certain limit corresponding to the noise we would like to measure whereas the relative shot noise keeps dropping. The main challenge was to maintain sufficient photodiode bandwidth without saturation, which is more likely for small photodiodes (0.5 mm diameter), while larger photodiodes (3 mm diameter) are bandwidth-limited to 60 MHz. Custom detection electronics were developed in which the photodiode was directly connected to a load resistor and the AC signal was extracted from the resulting voltage. Even at high optical powers saturating the photodiode, an AC signal could still be extracted. The photodetector frequency response was measured by using a characterized pig-tailed amplitude electro-optic modulator. 
\end{itemize}

The measurement results presented in Fig~.\ref{figure_RIN} show that:
\begin{itemize}
    \item At low frequencies (below $\approx 50$ Hz), the amplifier RPN was limited by turbulence effects and possible coupling of beam pointing noise through photodiode non-uniformity, but still remains below the AdV+ specification. 
    \item Between 30 Hz and 2 kHz, several peaks related to switch-mode power supplies of the NPRO (seeder) are observed. One can see that they are greatly reduced by the amplifier due to the amplification \cite{Novak:02} and that the remaining peaks are from the noise of the amplifier itself. Similarly, one also observes that from 50 - 500 Hz the noise after the fiber amplifier is slightly reduced with respect to the seeder noise.
    \item  The RPN transfer function of the whole amplifier was measured and used to project the seeder noise. This projection (in yellow) shows that the seeder contribution to the RPN of the MOPA output beam is negligible below 100 kHz. Above 100 kHz the amplifier does not produce any additional noise, and the output beam is strictly limited by the seeder noise at this level.
    \item Remarkably, above 30 MHz, the amplifier does not add any significant additional noise whereas one would have expected a contribution if the amplifier has significant SBS.
    \item For the whole frequency band of interest, the fiber amplifier complies with the AdV+ requirements. 
\end{itemize}

\subsection{Beam pointing noise}

\begin{figure}[hb!]
    \centering
    \includegraphics[width=1\linewidth]{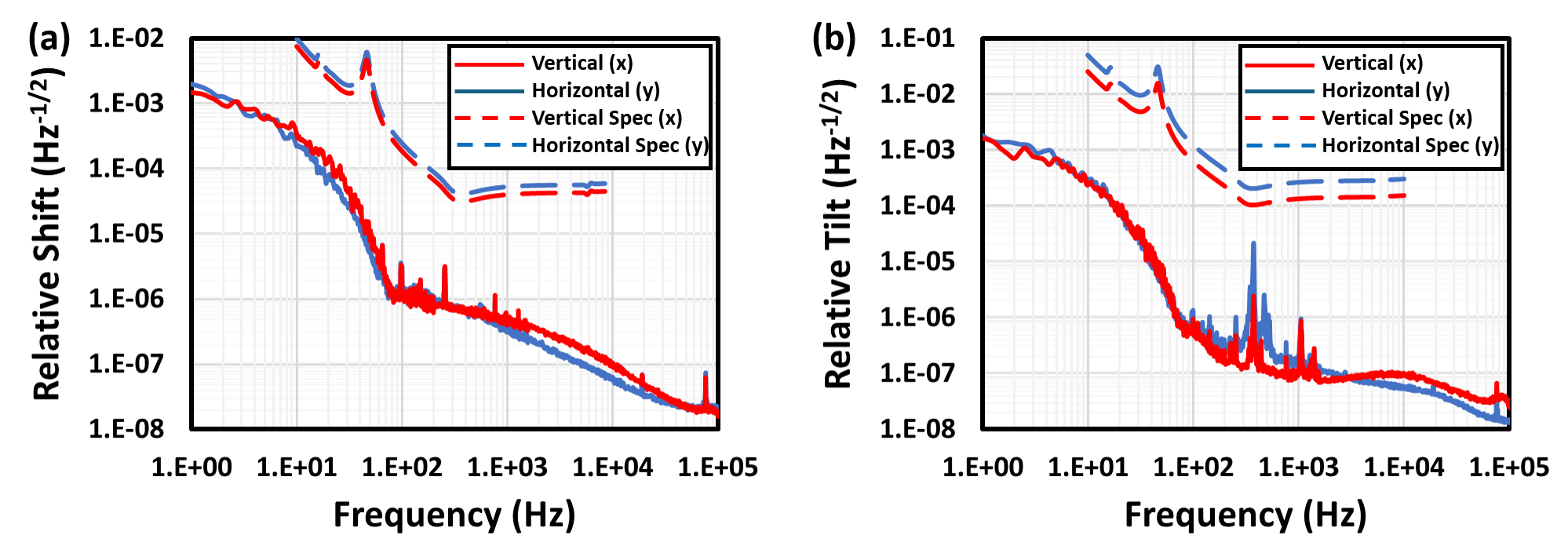}


    \includegraphics[width=0.7\linewidth]{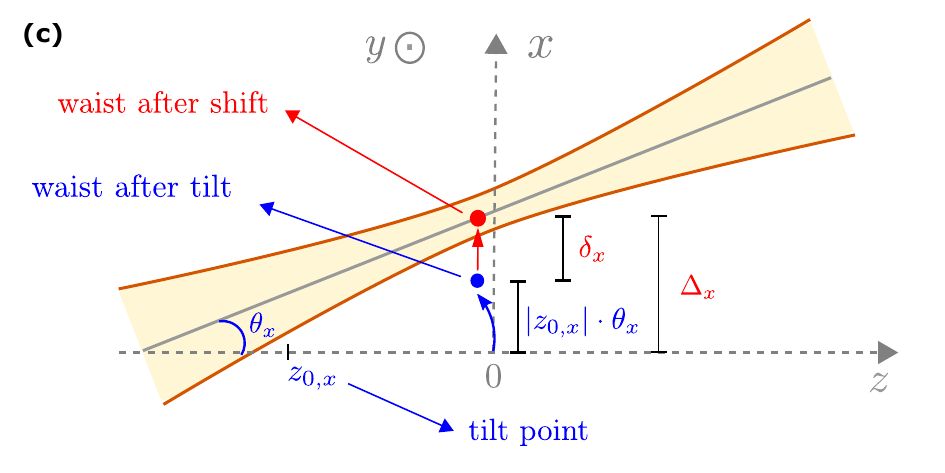}

    \caption{ASD of the beam pointing noise of the high power amplifier output beam separated into (a) shift noise and (b) tilt noise in both vertical and horizontal directions. Both measurements show a low frequency bump up to 100 Hz related to air turbulence effects. The measured noises remain below the Advanced Virgo specifications (dashed curves). (c) Representation of a beam jitter. Details in the text.}
    \label{figure_tilt_shift}
\end{figure}

Beam pointing noise is a particular concern in GW detectors. Considerable effort is devoted to isolate the test masses from seismic disturbances, resulting in highly stable and well-defined arm cavity modes within the detection band. The laser beam inevitably exhibits jitter, notably because the laser system itself is not seismically isolated. As will be described in Sec.~\ref{virgo_injection_system}, an input mode cleaner triangular cavity is present in the ISYS system with one of its purposes to filter beam pointing noise. Residual pointing fluctuations at its output lead to power fluctuations in the interferometer due to imperfect cavity coupling, which sets a requirement for the initial laser beam pointing noise.
The measurement of the beam pointing noise requires consideration of a reference optical axis with respect to which the laser beam is jittering, which could be an optical cavity as used in \cite{Kwee2007}. 
Let us assume that the laser beam is propagating along the $z$ axis and assume that it is in a pure Gaussian spatial mode with waist radius $w_0$ at $z_0=0$. Let us also assume that it is jittering around the $y$ axis, on the $xz$ plane, as shown in Fig.~\ref{figure_tilt_shift}c. Here the original beam with waist at $z_0 = 0$ and optical axis coincident with the $z$ axis was first tilted around the tilt point $z_{0,x}$ with an angle $\theta_x$, resulting in the new waist position marked by the blue circle. Subsequently, the beam was purely shifted along the $x$ axis by $\delta_x$, resulting in the final waist position indicated by the purple circle. 

To first order, the wave function that includes the jitter can be written as:
\begin{equation}
    \Psi\left(x,y,z,t\right)= \Psi_{0,0}\left(x,y,z\right) + \epsilon_x\left(t\right) \times \Psi_{1,0}\left(x,y,z\right) ,
\end{equation}where $\Psi_{0,0}\left(x,y,z\right)$ is the unperturbed fundamental Gaussian mode considered as reference, $\Psi_{1,0}\left(x,y,z\right)$ is the first higher order Hermite-Gauss mode~\cite{Siegman1986} along the $x$ axis of the same set of eigenmodes. 
With some algebra, we obtain (to first order):
\begin{eqnarray}
    \epsilon_x\left(t\right)&=&\left(\frac{\delta_x\left(t\right)}{w_0}-\frac{\theta_x\left(t\right)}{\theta_d}\times\frac{z_{0,x}\left(t\right)}{z_R}\right)-  \mathrm{i}\times\frac{\theta_x\left(t\right)}{\theta_d}\\
    &=&\frac{\Delta_x}{w_0}-\mathrm{i}\frac{\theta_x}{\theta_d},
\end{eqnarray}
where $\theta_d$ is the Gaussian beam divergence angle and $z_R$ is its Rayleigh range. The tilt term appears with a minus sign and contributes to an additional geometrical displacement of the beam optical axis at the beam waist position, as expressed in the generalized shift term $\Delta_x=\delta_x-\theta_x\times z_{0,x}$. Note that the given equations are an approximation to first order, where the beam shift along the $z$ axis is neglected. In this approximation, only two parameters are sufficient to characterize the jitter, instead of three, which in our case was chosen to be the tilt and the generalized shift. The amplitude $\left|\epsilon_x\left(t\right)\right|$ remains unchanged while propagating through a lens since, in the paraxial approximation, it doesn't couple the fundamental mode to higher-order mode. Hence, a matching telescope which couples the beam waist at the output of a laser system to the beam waist of a non degenerate cavity maintains the relative phase between the generalized shift and the tilt terms. Moreover, the beam shift projection to the cavity mode is given by a combination of the generalized shift term $\Delta_x$ and the tilt $\theta_x$ on which the Virgo specifications applies.
 
 The specific case of a lens placed at its focal length from the beam waist gives a beam waist image at the focal length further and adds a $\pi/2$ phase to $\epsilon_x$. This allows independent measurements of the beam tilt and generalized shift. 
  
The measurement of both terms was based on a split photodiode. Signals were obtained by taking the difference of the two voltages provided by each half photodiode after calibration. Only the in-phase term in $\epsilon$ with respect to the fundamental mode gives a linear contribution, the $\pi/2$ out of phase term gives a quadratic term which is then neglected. Therefore, the generalized shift is measured by placing the split photodiode in the beam waist right after the laser output, whereas the tilt is measured at the beam waist after the lens.
Let us first consider the measurement of the relative generalized shift $\Delta_{x,r}=\Delta_{x}/w_0$. The voltage difference is given by:
\begin{equation}
    \Delta V\left(t\right) = \alpha w_0 \left(\Delta_{x,r}\left(t\right)+\frac{\theta_x\left(t\right)}{\theta_d}\frac{\Delta z}{z_R}\right)+\alpha w_0\,\beta\, K\,\cdot \text{RPN}\left(t\right) ,
\end{equation}where:
\begin{itemize}
    \item $\alpha$ is the calibration factor in V/m obtained by measuring the mean voltage difference $\Delta V_0$ obtained by moving the split photodiode away from the center by a known distance $d_x$, i.e. $\alpha=\Delta V_0/d_x$.
    \item $\Delta z$ is the error on the longitudinal position of the photodiode with respect to the beam waist.
    \item $K$ stands for the difference of the sensitivities of the two half photodiodes, which is responsible for the coupling of the RPN into this measurement. It is usually $K \ll 1$.
    \item $\beta=\frac{1}{2}\sqrt{\frac{\pi}{2}}\simeq 0.63$ is a geometrical factor related to the Gaussian shape of the beam.
\end{itemize}In practice $z_R\simeq 1\,\textrm{m}$ and the position of the beam waist was determined within a couple of centimeters. Moreover, as shown in Figs.\ref{figure_tilt_shift}a and~\ref{figure_tilt_shift}b, $\theta/\theta_d\lesssim\Delta_{x,y}/w_0$. Hence, the $\frac{\theta_x\left(t\right)}{\theta_d}\frac{\Delta z}{z_R}$ could be neglected with respect to $\Delta_x$. Also, given the RPN level shown in Fig.~\ref{figure_RIN} and the obtained results on beam pointing noise shown in Fig.~\ref{figure_tilt_shift}, the last term can be neglected too. Ultimately, the generalized relative shift's spectrum is related to the voltage spectrum following:
\begin{equation}
    \Delta_{x,r}\left(f\right)\simeq \frac{\Delta V\left(f\right)}{\alpha w_0} ,
\end{equation}which is represented by the blue curve in Fig.~\ref{figure_tilt_shift}a.

The same procedure is followed to measure the tilt. Let $w'_0$ be the waist radius at the image focal plane of the lens (all physical quantities taken after the lens are marked with a prime). With some algebra, we obtain for the relative tilt $\Theta_{x,r}=\theta_x/\theta_d$:

\begin{equation}
    \Theta_{x,r}\left(f\right)\simeq \frac{\Delta V'\left(f\right)}{\alpha w'_0} . 
\end{equation}The relative tilt is displayed in blue in Fig.\ref{figure_tilt_shift}b. The same measurements are identically  performed in the $y$ direction and are displayed in red. These measurements are much lower than the requirement curves displayed by the dashed lines in Fig.\ref{figure_tilt_shift}, and hence are widely within the Virgo specifications. Elsewhere, the actual beam relative shift and tilt considered at the input mode cleaner cavity (see Sec.\ref{virgo_injection_system})  input are a combination of the measured $\Delta_{x\left(y\right),r}$ and $\Theta_{x\left(y\right),r}$ which are filtered by the PMC and the beam pointing control system~\cite{Canuel2014}.

\subsection{Long term behavior}
\subsubsection{Power fluctuation}
Since an essential prerequisite for laser systems in GW detectors are their long-term reliability, the fiber amplifier was tested for a total duration of 1886 hours ($\simeq2.5\,\textrm{months}$) before installation. This duration was adequate taking into account the following considerations:
\begin{itemize}
    \item After installation, the system was still monitored during the commissioning period. 
    \item The system was fully monitored during the test. In particular, the temperatures of the key elements (MFA, combiner, pumps diodes) were constantly measured in order to detect any precursor of a possible failure. 
    \item A decay of 1W/month, i.e. 12W/year, is acceptable because of the power margin. In this case, a two months test is able to identify such a decay, given the measurement accuracy of 1\%.
    \item In case of any sudden failure after installation, a spare system based on solid state lasers technology was available on the laser bench and only minutes are required to switch from one laser system to the other. A maintenance contract was also negotiated with the supplier to provide a rapid intervention. 
\end{itemize}

\begin{figure}[ht!]
\centering\includegraphics[width=\textwidth]{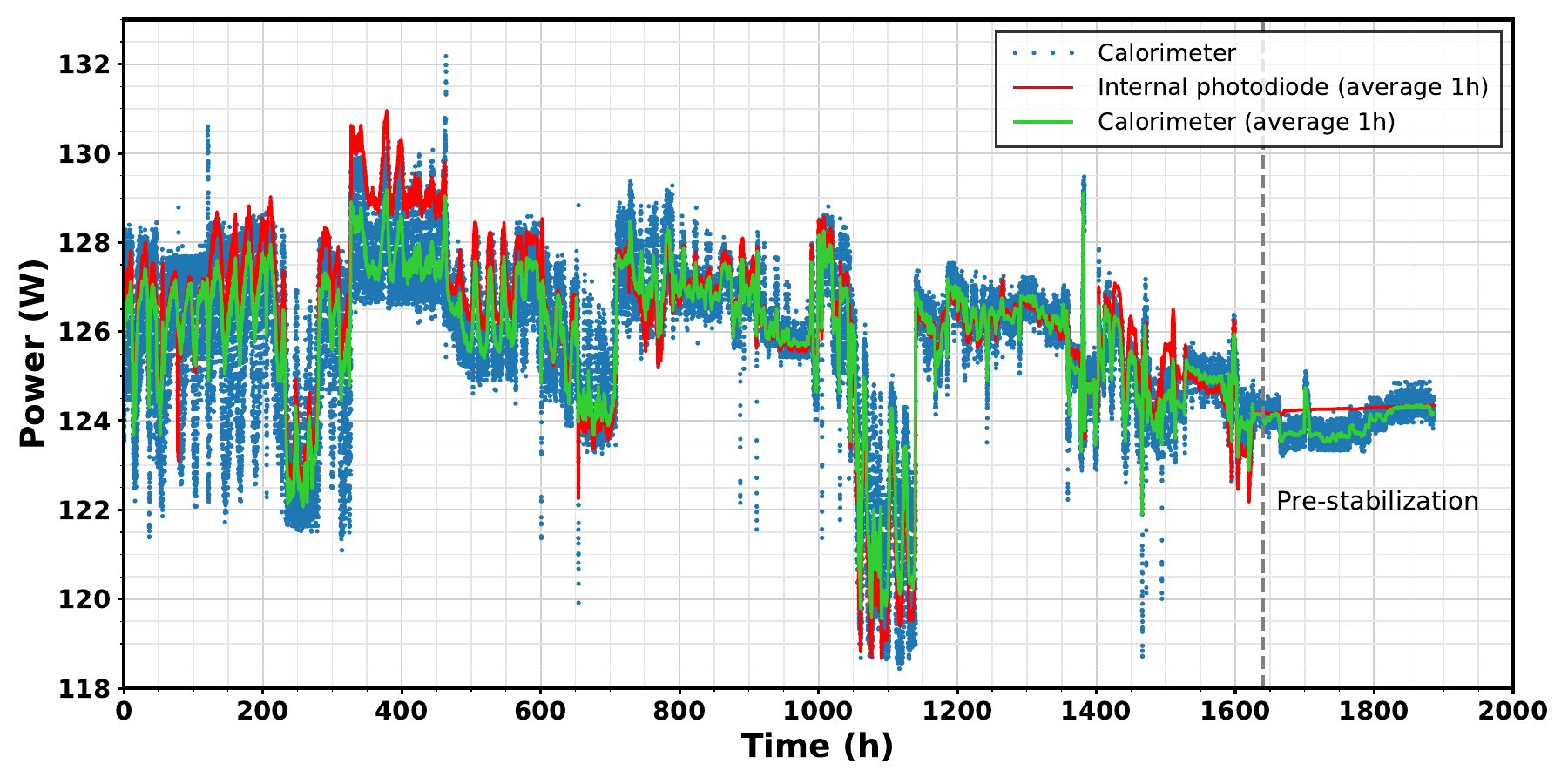}
\caption{Long term behavior of the output power for over 1800 hours during the validation test phase. Power measurements were acquired at one minute rate. The signal provided by a high power calorimeter (blue dots) and its average over 1 hour (green) are compared with the averaged signal from the internal photodiode inside the laser head (red). For the last 250 hours the laser was stabilized on the internal photodiode.}
\label{power_trend}
\end{figure}
The fiber amplifier was tested at an average power of approximately 125 W, 5 W of margin with respect to the maximum power (130 W) was taken for a long term ($>1$s) power pre-stabilization which was enganged for the last 250 hours of measurements. The output power was measured by a high power calorimeter characterized by an accuracy of $\pm 1\%$ and by a monitoring photodiode located inside the laser head that was calibrated with the calorimeter itself. Data was acquired with a sampling rate of $\textrm{minute}^{-1}$. 
Point to point discrepancies between the two signals were observed (not shown) but the moving average over one hour highlights the correlations between the two (see green and red curves in Fig.~\ref{power_trend}). The remaining differences are not significant because they are within the measurement accuracy (i.e. $<1$\%).  
The peak to peak power fluctuations reached 14 W over the measurement period (see blue dots in Fig.~\ref{power_trend}) and an apparent oscillation, corresponding to a period of 24 hours, is present. 
The pre-stabilization loop was based on the internal photodiode signal and the corrections were sent to the driving current of the laser diodes pumping the high-power amplifier. In this way, the long term peak to peak fluctuations on the calorimeter signal (out-of-loop) were reduced below 1 W.

\begin{figure}[ht!]
\centering\includegraphics[width=12cm]{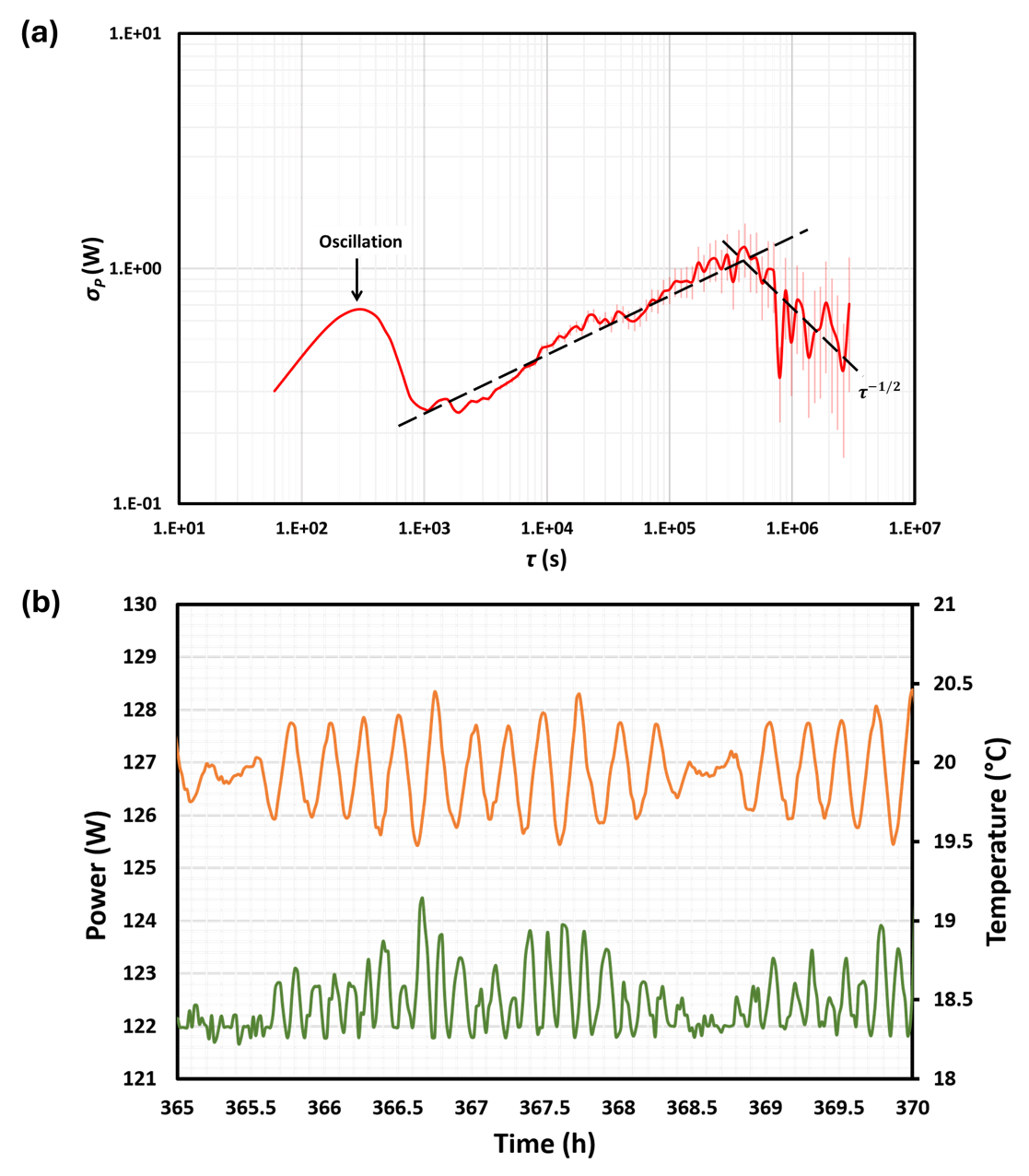}
\caption{(a) Allan deviation of the fiber amplifier output power showing (i) an oscillation with a 300 s period, (ii) a drift up to 6700 hours and (iii) a white noise for higher characteristic times. Vertical lines correspond to error bars. (b) 300 s power oscillation (in green) observed on 5 hours time scale related to the room temperature oscillation (in orange).}
\label{trend_allan_dev_trend_T}
\end{figure}

In an attempt to identify the origin of the observed fluctuations, the Allan standard deviation was computed and is displayed in Fig.~\ref{trend_allan_dev_trend_T} (a). We clearly identify three specific time ranges showing different behaviors:
\begin{itemize}
    \item an oscillation at 300 s;
    \item a drift for $10^3 \textrm{s}<\tau<4.10^5 \textrm{s}$ according to a law $\tau^\alpha$ with $\alpha\simeq 1/4$;
    \item a white noise integration following a law $\tau^{-1/2}$ for $\tau>4.10^5 \textrm{s}\simeq 111$ hours.
\end{itemize}
The oscillation at 300 s was likely related to the room temperature as shown in Fig.\ref{trend_allan_dev_trend_T}b. The amplitude of the power fluctuations depended on the room temperature fluctuations due to the air conditioning system, which are of $1^\circ$ peak to peak. In addition, the power oscillations appeared at twice the frequency of the room temperature fluctuations. This shows that at the room temperature of 20$^\circ$, the system was in an optimal working point for which the dependence of the output power with respect to the room temperature was quadratic. According to the manufacturer, this coupling originates from temperature-dependent polarization fluctuations in the fiber that transports the pre-amplified beam to the high-power amplifier (see Fig.~\ref{figure_amplifier_architecture}). For integration reasons, this fiber was specified to have a length of 2 m. Up to $5\times10^5$s, the range over which the power varies is mainly due to the drift $\simeq\tau^{1/4}$. The daily modulation visible in Fig.~\ref{power_trend} is probably hidden in the drift.
Large peak to peak fluctuations of $\simeq 2^\circ$ C were observed on the room temperature but no correlations could be established with largest power fluctuations occured for example at $t = 1050\,\text{h}$.

\subsubsection{Beam pointing drift}

\begin{figure}[ht!]
\centering\includegraphics[width=12cm]{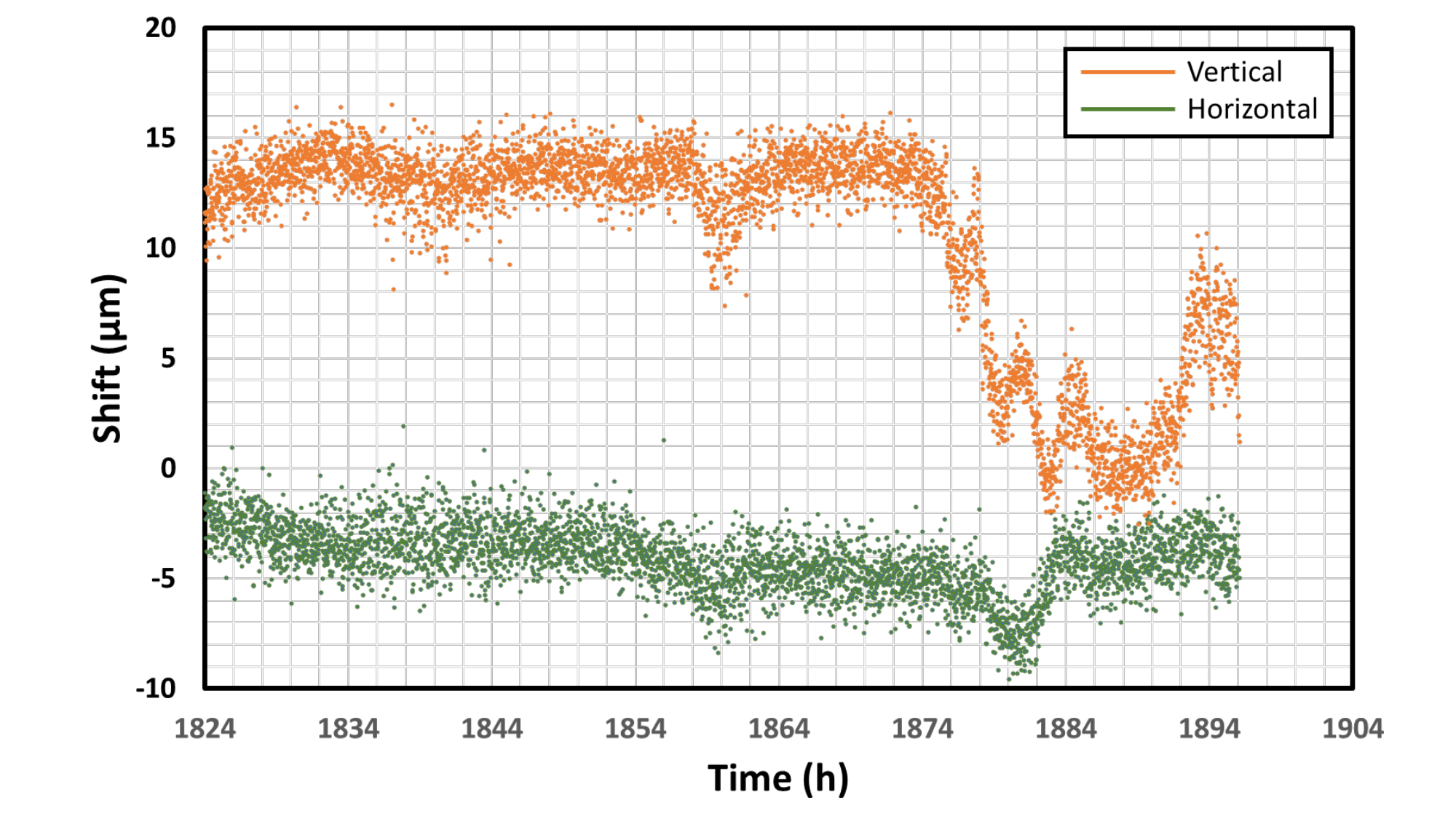}
\caption{Vertical and horizontal beam pointing drift measured over three days. 
}
\label{Drift}
\end{figure}

Apart from the intrinsic power fluctuations of the amplified laser beam, another source of power fluctuation/drift had to be considered: when the laser beam is injected into a mode cleaner cavity, the beam pointing drift is transformed into power drift at the cavity output. 
To characterize this, a measurement of the beam drift was performed using a quadrant photodiode placed at the fiber amplifier output beam waist with a minimum number of optics. Both vertical and horizontal shift are displayed in Fig.~\ref{Drift} over roughly 100 hours, which corresponds the characteristic time of the maximum power drift; see Fig~\ref{trend_allan_dev_trend_T}a. The plots show that the vertical drift is dominant, which might be related to the cooling of the laser head which is performed by its lower surface. The maximum drift $\delta$ was of the order of $20~\mu\textrm{m}$, which corresponds to a relative power drift through a mode cleaner cavity $\delta P/P\simeq\left(\delta/w_0\right)^2<0.2\%$ that is negligible with respect to the intrinsic power drift.

\section{On site system operation}

In this section we describe the installation and commissioning of the fiber amplifier with the Virgo injection system. Careful planning is required to coordinate the installation with the detector infrastructure and ongoing activities. The commissioning process is naturally subject to several constraints, as the integration and characterization of the amplifier depends on the operation and configuration of the entire detector.

\subsection{The Virgo injection system}
\label{virgo_injection_system}

\begin{figure}[ht!]
\centering\includegraphics[trim= 0cm 0cm 0cm 0cm,clip,width=12cm]{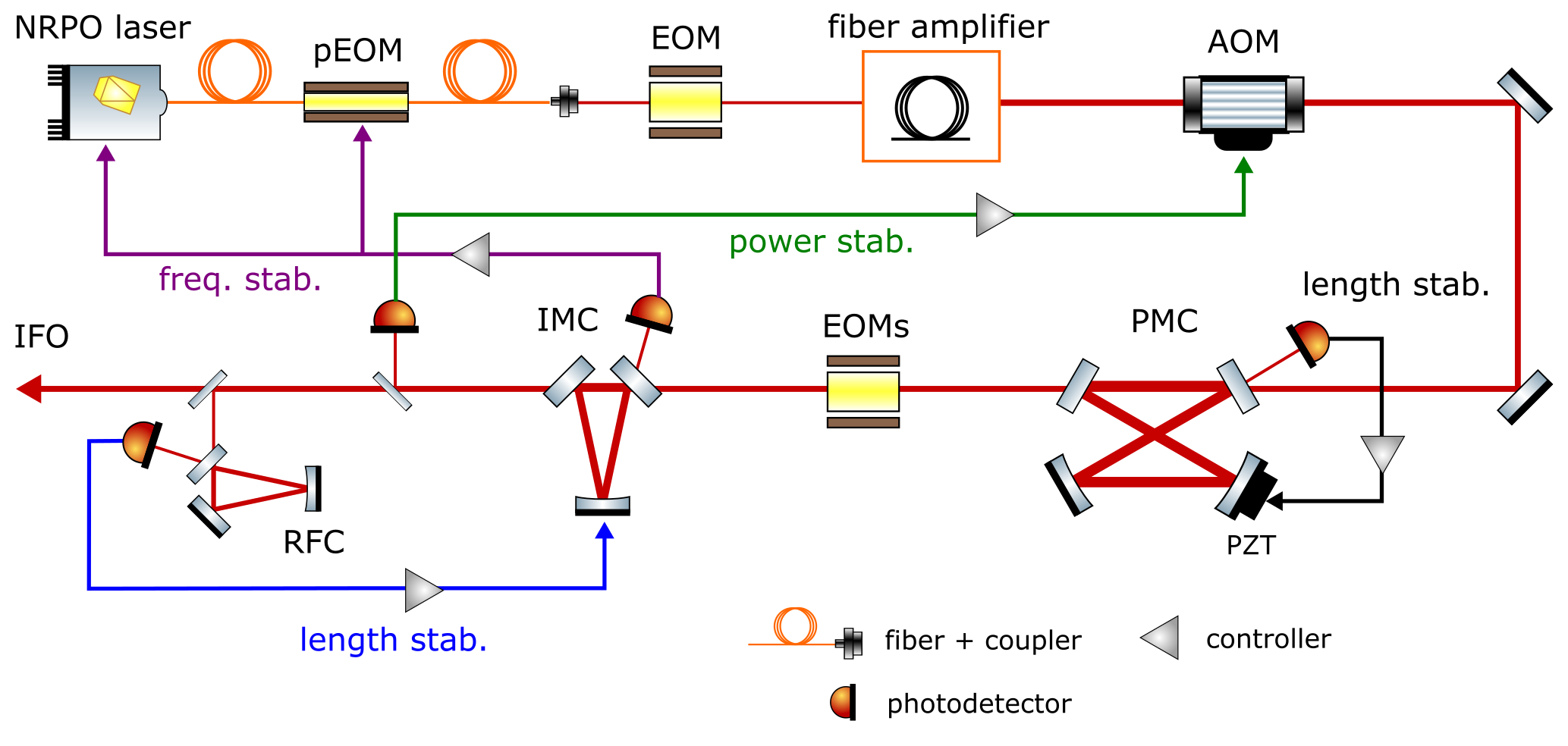}
\caption{Simplified layout of the injection system including the pre-stabilized laser which includes three stabilization loop : (i) the laser frequency is stabilized on the IMC (ii) the IMC transmitted power is stabilized on a set of two photodiodes (iii) the PMC is stabilized on the laser frequency. The IMC length is stabilized at low frequency on the RFC.}
\label{Injection_system}
\end{figure}
Fig.~\ref{Injection_system} shows a simplified layout of the injection system (ISYS) of Virgo, quickly introduced in Sec. \ref{sec:PSL}. The main components of the system are a laser source, a fiber amplifier, and three cavities: the pre-mode cleaner cavity (PMC), the input mode cleaner cavity (IMC), and the reference cavity (RFC). The low power laser source is a 1 W Mephisto Model from Coherent$^\circledR$ which is a single frequency and fiber coupled NPRO. Its output beam was injected into a pigtailed Electro-Optic Modulator (pEOM), used as a frequency actuator (see next section). After exiting a fiber through a fiber collimator, it propagates in free space on the laser bench (red traces). The PMC is a bow tie shape cavity with a round trip of 1.6 m and a Finesse of 120 corresponding to a cavity pole of 780 kHz. Its length is stabilized to the laser frequency via a PDH scheme~\cite{Drever1983}, with sidebands at $\sim 14~\textrm{MHz}$ generated by the free space Electro-Optic Modulator (EOM) before the fiber amplifier. The main role of the PMC is to spatially filter the laser mode and reduce beam pointing noise. After the PMC, the beam travels through a set of homemade EOMs to generate several side-bands for locking and control purposes. The IMC is a 144 m long triangular and suspended cavity with a Finesse of 1000 which corresponds to a cavity pole of 500 Hz. Sidebands at $\sim 22~\textrm{MHz}$ are reflected by the IMC cavity and are used to stabilize the laser frequency to the IMC length.  Its transmitted beam is sent to the interferometer (IFO). Like the PMC, the IMC also filters the beam shape and pointing noise, and additionally it acts as a frequency reference for the laser frequency stabilization. The schematics also shows a power stabilization loop, which senses a sample of the main beam with two photodetectors (not shown in the figure for simplicity) and sends the correction signal to an Acousto-Optic Modulator (AOM) placed before the PMC. The AOM acts on the laser power by transferring optical power between the zero-order main beam and the first-order diffracted beam. A second beam is sampled from the main beam and injected to an ultra stable RFC which is used as a reference to stabilize the IMC length at low frequencies (up to 200 Hz). The RFC has been already described in Sec.\ref{Section_frequency_noise} and its frequency pole is $\sim 250~\textrm{kHz}$. 

The fiber amplifier was installed and commissioned at a power corresponding to 40 W transmitted by the IMC and sent to the interferometer, which was larger than the required power of 25 W. 

In the following, we present the performance of the stabilization loops within the injection system after the installation of the high-power fiber amplifier. The objective was to demonstrate that the fiber amplifier does not constitute a limiting factor for the operation of the injection system with respect to the specification required by the interferometer.

\subsection{Performance of the Control Loops}
\subsubsection{Frequency pre-stabilization loop}

\begin{figure}[ht!]
\centering\includegraphics[trim= 0cm 0cm 0cm 0cm,clip,width=12cm]{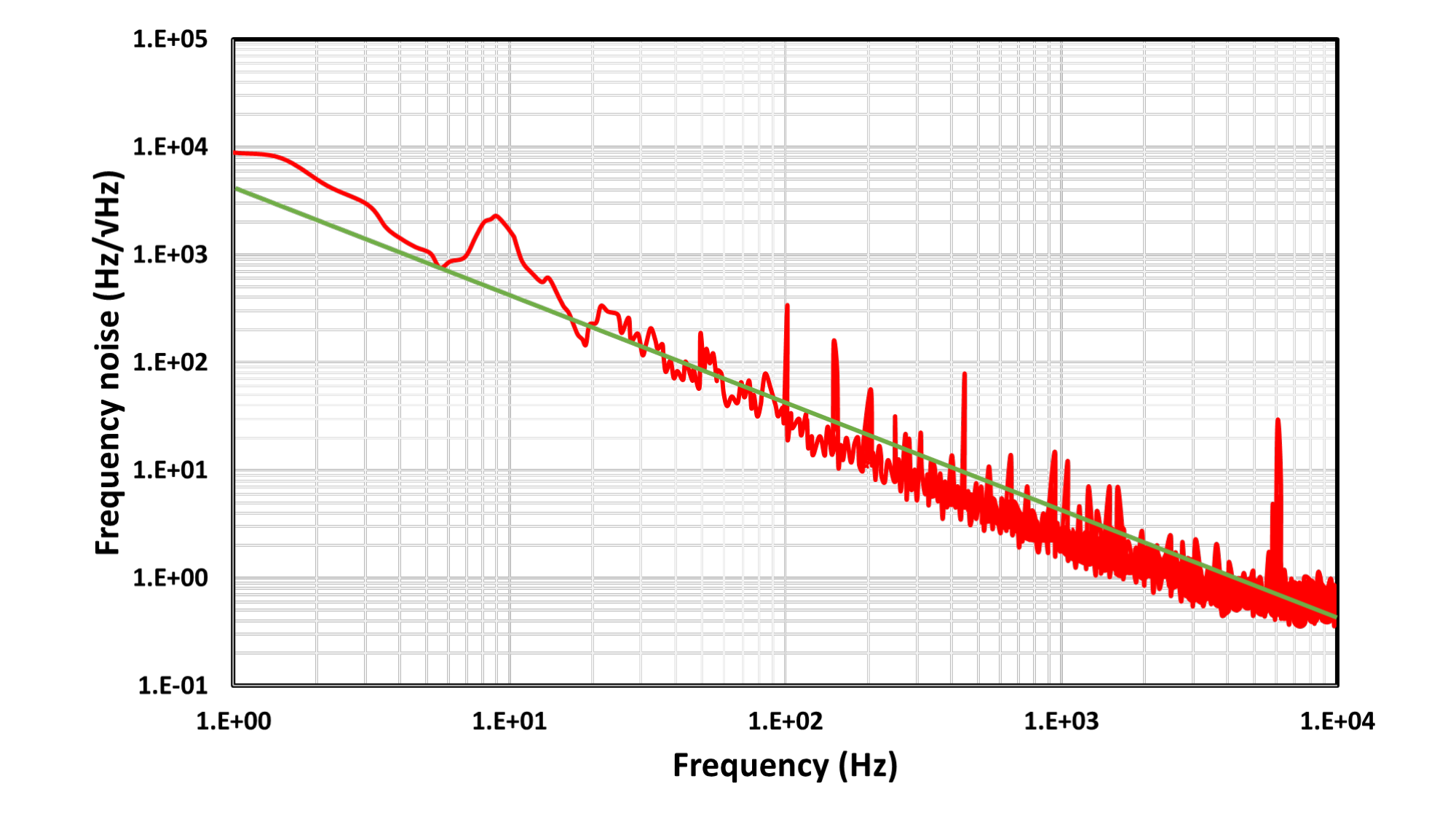}
\caption{Free running frequency noise of the laser measured from the in-loop correction signal sent to the PZT actuator of the NPRO (red). The green line corresponds to the $1/f$ expected NRPO behavior fitted to the data.}
\label{IMC_PZT}
\end{figure}

The frequency of the laser beam at the interferometer input is first pre-stabilized to the length of the IMC and then further stabilized to the common mode of the Michelson interferometer. This latter control loop is referred to as the Second Stage Frequency Stabilization (SSFS)~\cite{VANDAEL2025}, and will not be considered further in this paper. The Unity Gain Frequency (UGF) of the pre-stabilization loop was approximately 100 kHz with a phase margin from 15 to 20 degrees~\cite{VANDAEL2025}. The actuation was three fold depending on the frequency range, and the corrections were applied to:
\begin{itemize}
    \item the NPRO crystal temperature for $f\lesssim1~\textrm{Hz}$,
    \item the NPRO PZT actuator for $1~\textrm{Hz}\lesssim f \lesssim 10~\textrm{kHz}$,
    \item the pigtailed EOM for $f\gtrsim 10~\textrm{kHz}$.
\end{itemize}

From the correction sent to the PZT actuator, we inferred the frequency noise depicted in the red curve of Fig.~\ref{IMC_PZT}. The curve shows that the amplified beam preserves the $1/f$ behavior typical of the NPRO (in green). This is a confirmation that the fiber amplifier did not add significant frequency noise, as already tested and shown in Fig.~\ref{figure_frequency_noise}. Note that the overall frequency noise is measured to be one fifth of the one measured during the validation tests which were realized with a different NPRO.

\begin{figure}[ht!]
\centering\includegraphics[trim= 0cm 0cm 0cm 0cm,clip,width=12cm]{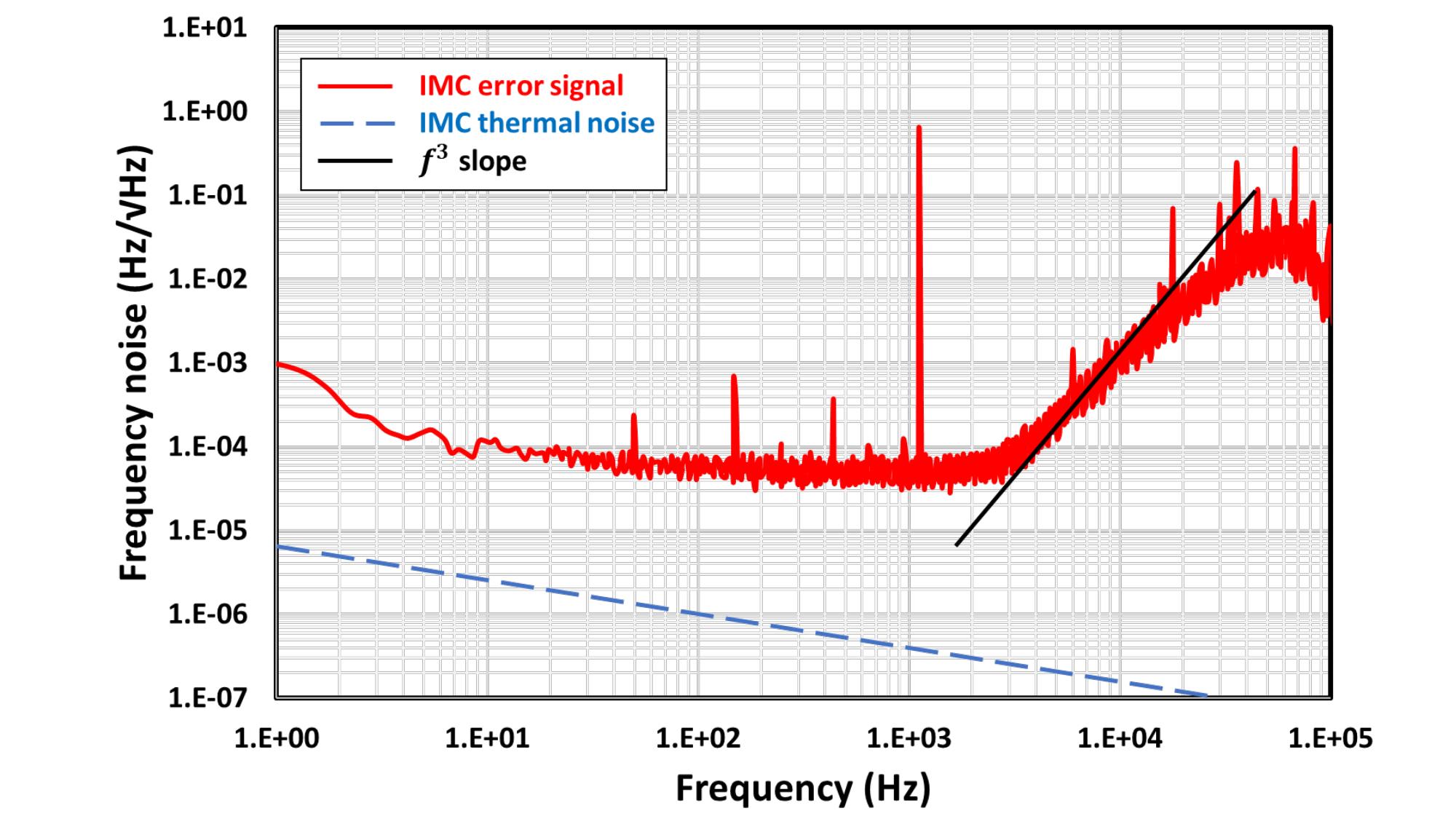}
\caption{ASD of the calibrated error signal of the frequency pre-stabilization loop into frequency noise (red cruve). Above 3 kHz, the loop was gain limited (black line). Between 10 Hz and 3 kHz, sensing noise limited the frequency stability to $5\times 10^{-5}\,\textrm{Hz}/\sqrt{\textrm{Hz}}$. The dashed-blue line shows an estimation of the frequency noise of the IMC due to mirror thermal noise limited stability. }
\label{IMC_error_signal}
\end{figure}

The in-loop frequency noise is shown in Fig.\ref{IMC_error_signal}. The open-loop transfer function (OLTF) of the IMC loop exhibited an integrator-like behavior, with a \(1/f\) dependence from the unity-gain frequency (UGF) at 100 kHz down to approximately 30 kHz, below which the gain was proportional to \(1/f^4\). From 3 kHz until $30~\textrm{kHz}$ the noise suppression was limited by the loop gain, as can be seen by the an error signal with a $f^3$ behavior (since the frequency noise is $\propto 1/f$). Below 3 kHz, the residual noise was approximately frequency independent down to 10 Hz and it was concluded to be limited by sensing noise at the level of $5\times10^{-5}~\textrm{Hz}/\sqrt{\textrm{Hz}}$, which is well above the IMC mirror thermal noise estimated in the dashed blue line. It was then concluded that the intrinsic noise of the fiber amplifier did not limit the frequency loop. The 1111 Hz line observed in the measurement was intentionally injected for control purposes. 

\subsubsection{Power stabilization loop}
\begin{figure}[ht!]
\centering\includegraphics[trim= 0cm 0cm 0cm 0cm,clip,width=12cm]{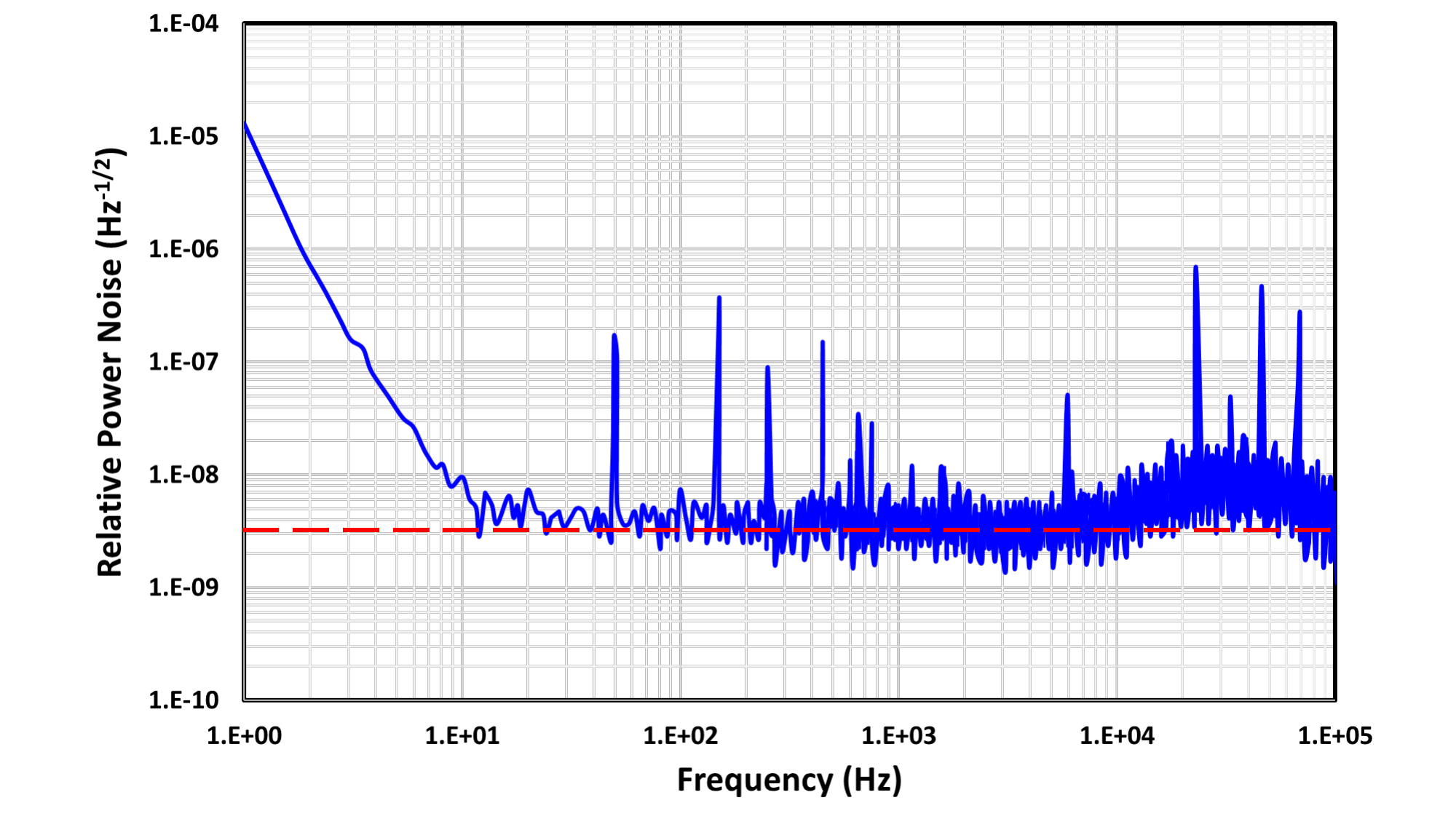}
\caption{ASD of the RPN measured by the out-of-loop photodiodes after the IMC (blue curve). The total uncorrelated shot noise of the in-loop and out-of-loop photodetectors is displayed by the dashed red line. Details in the text. }
\label{Pstab_RIN}
\end{figure}

The power stabilization loop (green line in Fig.~\ref{Injection_system}) was based on a classic scheme in which the power fluctuations of a pick-off beam are split and sensed by two in-loop photodiodes \cite{Kwee2009}. The stabilized beam was monitored using two out-of-loop photodiodes (not shown in the figure, but also placed in transmission of the IMC) which detected the same total power as the in-loop power. During the fiber amplifier operation, the total power sensed by both in-loop and out-of-loop photodiodes was 70 mW corresponding to approximately 60 mA of total photocurrent. The power stabilization loop acted on frequencies higher than 0.1 Hz and had a UGF of 100 kHz. As seen in Fig.~\ref{Pstab_RIN}, the RPN measured by the out-of-loop photodiodes reached a noise floor of $ \simeq 3.2\times 10^{-9} \, \textrm{Hz}^{-1/2}$ between 50 Hz and 10 kHz corresponding to $\sqrt{2}$ times the relative shot noise level of $\simeq 2.3 \times 10^{-9} \, \textrm{Hz}^{-1/2}$ at the measured total power by the in-loop photodiodes. The $\sqrt{2}$ factor corresponds to the additional uncorrelated shot noise sensed by the out-of-loop photodiodes. Between 10 Hz and 50 Hz, the RPN increased up to $3.5\times 10^{-9}~\textrm{Hz}^{-1/2}$ due to the photodiodes dark noise. Since the relative shot noise of the main beam sent to the interferometer at 40 W is negligible, its residual RPN was estimated to be $\simeq 2.5\times 10^{-9}~\textrm{Hz}^{-1/2}$ in the 10 Hz - 50 Hz band and $\simeq 2.3\times 10^{-9}~\textrm{Hz}^{-1/2}$ in the 50 Hz - 10 kHz band which is below the interferometer requirements. This was further verified through a noise injection, which showed that the RPN at the interferometer input was at a level more than a factor of 10 below the target interferometer sensitivity. Below 10 Hz, the power stability was gain limited. 

\subsubsection{The PMC length stabilization loop}
\begin{figure}[ht!]
\centering\includegraphics[trim= 0cm 0cm 0cm 0cm,clip,width=12cm]{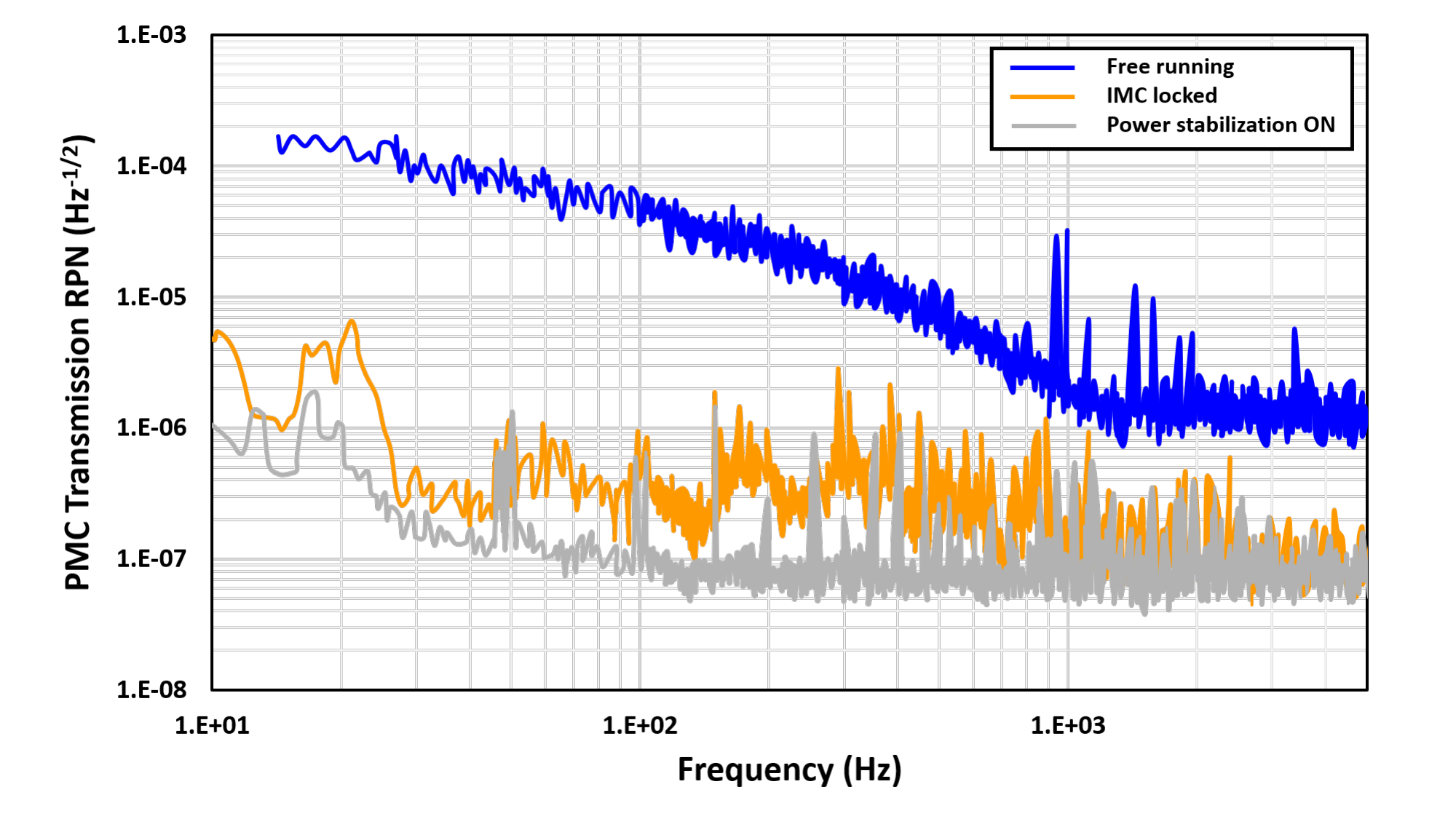}
\caption{ASD of the RPN of the PMC transmitted beam for three different cases: with both the power stabilization and frequency pre-stabilization loops disengaged (blue); (ii) with only the frequency pre-stabilization loop to the IMC engaged (orange); and (iii) with both the power stabilization and frequency pre-stabilization loops engaged (grey). Ultimately, the RPN was limited by beam pointing noise coupled to the PMC. The shot noise limit lies at $4.3\times 10^{-9}\text{Hz}^{-1/2}$ and is not represented here.}
\label{PMC_TRA}
\end{figure}

The PMC length was stabilized to the frequency of the high power beam with a coupling efficiency of 89\%. The UGF of the PMC loop was 4 kHz. We characterized the loop behavior by measuring the power noise of the transmitted beam. In Fig.~\ref{PMC_TRA}, the RPN of the beam transmitted through the PMC is presented for three configurations: (i) with both the power stabilization and frequency pre-stabilization loops disengaged (blue curve); (ii) with only the frequency pre-stabilization loop to the IMC engaged (orange curve); and (iii) with both the power stabilization and frequency pre-stabilization loops engaged. The reduction in the transmitted power noise after the IMC loop was engaged shows a coupling between frequency noise to power noise through the PMC, and it is a signature of a small detuning with respect to the PMC resonance. By engaging the power stabilization loop, the power noise was reduced and is approximately frequency independent after 100 Hz. Since the measured beam shot noise level was at $\simeq 4.3\times 10^{-9}\,\text{Hz}^{-1/2}$, we suspect that the noise floor is due to the residual beam pointing noise of the laser with respect to the IMC, which is injected by the Pstab loop at the PMC level.

\subsection{Long term behaviour}

\begin{figure}[ht!]
\centering\includegraphics[trim= 0cm 0cm 0cm 0cm,clip,width=12cm]{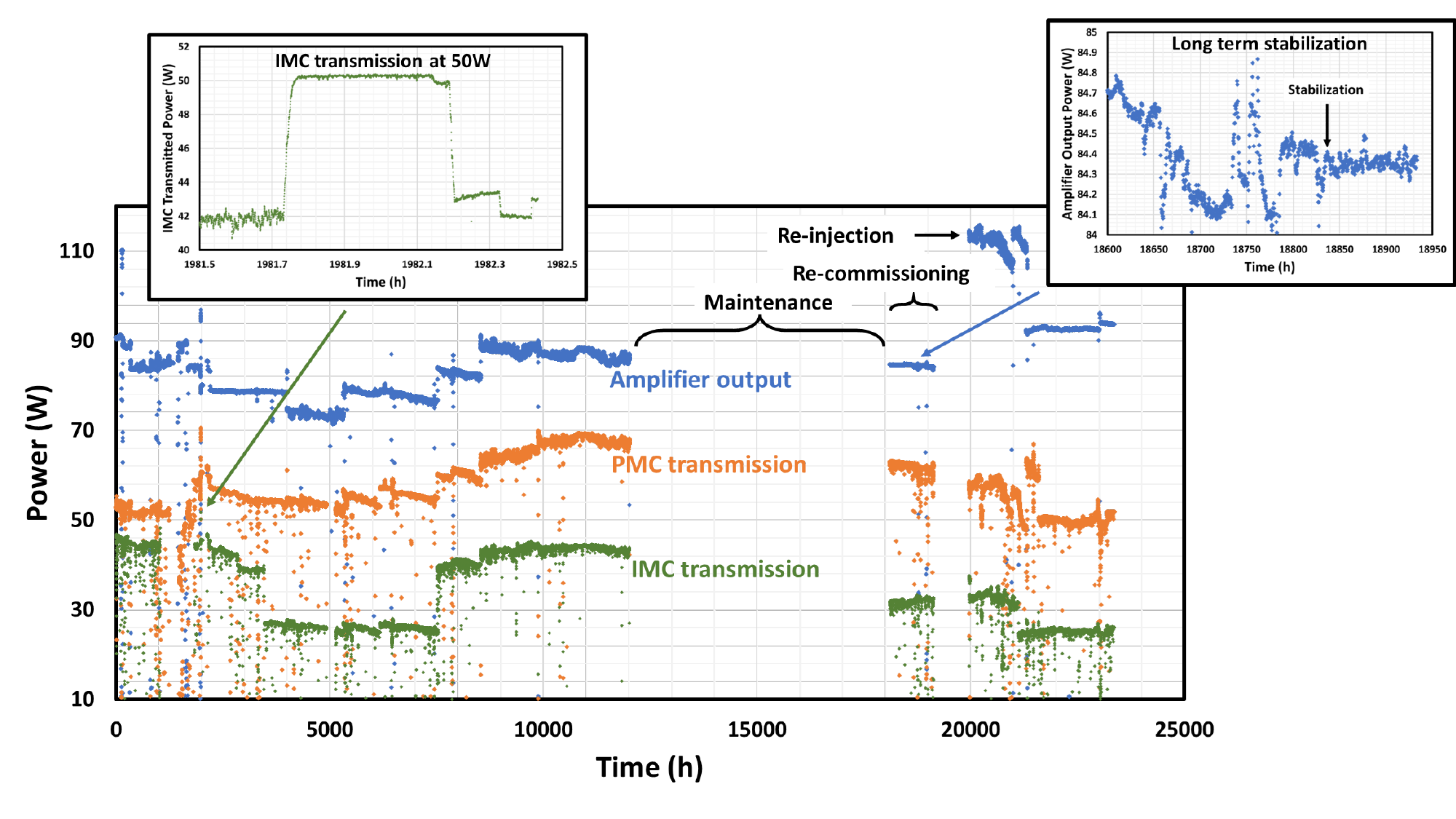}
\caption{Long term behavior of the power at the fiber amplifier output (blue), in transmission of PMC (orange), and in transmission of the IMC (green). Left inset: the fiber amplifier power was increased such that the IMC transmission reached 50 W. Right inset: a long term power stabilization loop was implemented to reduce the power fluctuations of the laser beam sent to the interferometer.}
\label{Long_term_EGO}
\end{figure}

For over 17000 hours, the high power fiber amplifier has been used for the commissioning of the Virgo interferometer and all the subsystems that required the pre-stabilized laser. It corresponded to a period ranging from October 2020 to March 2023 with a maintenance which lasted for 7 months. During commissioning, the power delivered to the interferometer was initially set to 40 W, then reduced to 30 W, and finally to 25 W (according to the commissioning requirements). The power provided by the injection system was also tested up to 50 W as shown on the left inset of Fig.~\ref{Long_term_EGO}. A long term stabilization system was implemented through the acquisition and control system of the whole detector. As shown on the right inset of Fig.~\ref{Long_term_EGO}, the peak-to-peak fluctuations were reduced to 200 mW over 100 hours for an output power 84 W. 

After March 2023 the fiber amplifier system was removed, prior to the O4 observing run~\cite{Arnaud2026}, due to the observation of an additional noise source in the interferometer sensitivity. The origin of this noise was identified through dedicated investigations to be the coupling of high frequency (MHz regime) phase noise at the input of the IMC. These issues, including the origin and coupling of this noise, could not be fully addressed in time for the fiber amplifier to be used during O4. The nature and characterization of this noise, as well as the associated investigations, will be presented in a forthcoming publication. 

\section{Discussion and Conclusion}

In this work, we have presented the first installation and commissioning of a high-power fiber amplifier in a gravitational-wave detector, demonstrating the delivery of a high-power, single-mode, and single-frequency laser beam compatible with the stringent requirements of gravitational-wave interferometry. The amplifier was an all-fiber system based on a large-mode-area ytterbium-doped fiber and delivered up to 130 W of optical power.

The system was first extensively characterized and validated at the Artemis laboratory. The modal content analysis showed that 92.6 \% of the output power was carried by the fundamental mode. Remarkably, the amplification process was found not to introduce additional frequency noise within the gravitational-wave detection bandwidth, up to 10 kHz. The relative power noise of the amplified beam remained below the requirements of the Advanced Virgo+ project. Furthermore, no additional power noise was observed between 200~ kHz and 100 MHz with respect to the seed laser, indicating the absence of stimulated Brillouin scattering in the amplifier. The system also fulfilled the requirements on beam-pointing noise and long-term stability. With a power pre-stabilization loop implemented, we measured a long-term peak-to-peak power fluctuations below 1 W.

Following these laboratory tests, the amplifier was successfully installed and commissioned at the Virgo detector within the framework of the Advanced Virgo+ project. The amplified laser beam was integrated into the frequency pre-stabilization system and stabilized in power to a level of approximately $2.3 \times 10^{-9}~\mathrm{Hz}^{-1/2}$. Interestingly, the free-running frequency noise measured at the Virgo site was approximately five times lower than that measured during the laboratory tests. 

The long-term performance of the amplifier represents another important result of this work. The system operated continuously for approximately two and a half months during the laboratory tests and for almost two years during the Virgo detector commissioning. The successful integration of the amplifier ultimately enabled an optical power of 50 W at the interferometer input.

The results of the implementation of the fiber amplifier at Virgo demonstrate that high-power fiber amplification is a promising technology for gravitational-wave detectors, and they were an important step towards their use in the next generation of detectors where substantially higher laser powers will be required.

\begin{backmatter}

\bmsection{Acknowledgments}
 The authors would like to thank Benno Willke and Matteo Tacca for their careful reading of the manuscript and for their valuable comments and suggestions. 

\bmsection{Disclosures}
The authors declare no conflicts of interest.

\bmsection{Data availability} Data underlying the results presented in this paper are not publicly available at this time but may be obtained from the authors upon reasonable request.

\bigskip

\end{backmatter}

\bibliography{sample}

@article{GWTC1,
  title = "{GWTC-1: A Gravitational-Wave Transient Catalog of Compact Binary Mergers Observed by LIGO and Virgo during the First and Second Observing Runs}",
  author = {Abbott, B. P. and Abbott, R. and Abbott, T. D. and Abraham, S. and Acernese, F. and Ackley, K. and Adams, C. and Adhikari, R. X. and Adya, V. B. and Affeldt, C. and Agathos, M. and Agatsuma, K. and Aggarwal, N. and Aguiar, O. D. and Aiello, L. and Ain, A. and Ajith, P. and Allen, G. and Allocca, A. and Aloy, M. A. and Altin, P. A. and Amato, A. and Ananyeva, A. and Anderson, S. B. and Anderson, W. G. and Angelova, S. V. and Antier, S. and Appert, S. and Arai, K. and Araya, M. C. and Areeda, J. S. and Ar\`ene, M. and Arnaud, N. and Arun, K. G. and Ascenzi, S. and Ashton, G. and Aston, S. M. and Astone, P. and Aubin, F. and Aufmuth, P. and AultONeal, K. and Austin, C. and Avendano, V. and Avila-Alvarez, A. and Babak, S. and Bacon, P. and Badaracco, F. and Bader, M. K. M. and Bae, S. and Baker, P. T. and Baldaccini, F. and Ballardin, G. and Ballmer, S. W. and Banagiri, S. and Barayoga, J. C. and Barclay, S. E. and Barish, B. C. and Barker, D. and Barkett, K. and Barnum, S. and Barone, F. and Barr, B. and Barsotti, L. and Barsuglia, M. and Barta, D. and Bartlett, J. and Bartos, I. and Bassiri, R. and Basti, A. and Bawaj, M. and Bayley, J. C. and Bazzan, M. and B\'ecsy, B. and Bejger, M. and Belahcene, I. and Bell, A. S. and Beniwal, D. and Berger, B. K. and Bergmann, G. and Bernuzzi, S. and Bero, J. J. and Berry, C. P. L. and Bersanetti, D. and Bertolini, A. and Betzwieser, J. and Bhandare, R. and Bidler, J. and Bilenko, I. A. and Bilgili, S. A. and Billingsley, G. and Birch, J. and Birney, I. A. and Birnholtz, O. and Biscans, S. and Biscoveanu, S. and Bisht, A. and Bitossi, M. and Bizouard, M. A. and Blackburn, J. K. and Blackman, J. and Blair, C. D. and Blair, D. G. and Blair, R. M. and Bloemen, S. and Bode, N. and Boer, M. and Boetzel, Y. and Bogaert, G. and Bondu, F. and Bonilla, E. and Bonnand, R. and Booker, P. and Boom, B. A. and Booth, C. D. and Bork, R. and Boschi, V. and Bose, S. and Bossie, K. and Bossilkov, V. and Bosveld, J. and Bouffanais, Y. and Bozzi, A. and Bradaschia, C. and Brady, P. R. and Bramley, A. and Branchesi, M. and Brau, J. E. and Briant, T. and Briggs, J. H. and Brighenti, F. and Brillet, A. and Brinkmann, M. and Brisson, V. and Brockill, P. and Brooks, A. F. and Brown, D. D. and Brunett, S. and Buikema, A. and Bulik, T. and Bulten, H. J. and Buonanno, A. and Buskulic, D. and Bustamante Rosell, M. J. and Buy, C. and Byer, R. L. and Cabero, M. and Cadonati, L. and Cagnoli, G. and Cahillane, C. and Calder\'on Bustillo, J. and Callister, T. A. and Calloni, E. and Camp, J. B. and Campbell, W. A. and Canepa, M. and Cannon, K. C. and Cao, H. and Cao, J. and Capocasa, E. and Carbognani, F. and Caride, S. and Carney, M. F. and Carullo, G. and Casanueva Diaz, J. and Casentini, C. and Caudill, S. and Cavagli\`a, M. and Cavalier, F. and Cavalieri, R. and Cella, G. and Cerd\'a-Dur\'an, P. and Cerretani, G. and Cesarini, E. and Chaibi, O. and Chakravarti, K. and Chamberlin, S. J. and Chan, M. and Chao, S. and Charlton, P. and Chase, E. A. and Chassande-Mottin, E. and Chatterjee, D. and Chaturvedi, M. and Chatziioannou, K. and Cheeseboro, B. D. and Chen, H. Y. and Chen, X. and Chen, Y. and Cheng, H.-P. and Cheong, C. K. and Chia, H. Y. and Chincarini, A. and Chiummo, A. and Cho, G. and Cho, H. S. and Cho, M. and Christensen, N. and Chu, Q. and Chua, S. and Chung, K. W. and Chung, S. and Ciani, G. and Ciobanu, A. A. and Ciolfi, R. and Cipriano, F. and Cirone, A. and Clara, F. and Clark, J. A. and Clearwater, P. and Cleva, F. and Cocchieri, C. and Coccia, E. and Cohadon, P.-F. and Cohen, D. and Colgan, R. and Colleoni, M. and Collette, C. G. and Collins, C. and Cominsky, L. R. and Constancio, M. and Conti, L. and Cooper, S. J. and Corban, P. and Corbitt, T. R. and Cordero-Carri\'on, I. and Corley, K. R. and Cornish, N. and Corsi, A. and Cortese, S. and Costa, C. A. and Cotesta, R. and Coughlin, M. W. and Coughlin, S. B. and Coulon, J.-P. and Countryman, S. T. and Couvares, P. and Covas, P. B. and Cowan, E. E. and Coward, D. M. and Cowart, M. J. and Coyne, D. C. and Coyne, R. and Creighton, J. D. E. and Creighton, T. D. and Cripe, J. and Croquette, M. and Crowder, S. G. and Cullen, T. J. and Cumming, A. and Cunningham, L. and Cuoco, E. and Canton, T. Dal and D\'alya, G. and Danilishin, S. L. and D'Antonio, S. and Danzmann, K. and Dasgupta, A. and Da Silva Costa, C. F. and Datrier, L. E. H. and Dattilo, V. and Dave, I. and Davier, M. and Davis, D. and Daw, E. J. and DeBra, D. and Deenadayalan, M. and Degallaix, J. and De Laurentis, M. and Del\'eglise, S. and Del Pozzo, W. and DeMarchi, L. M. and Demos, N. and Dent, T. and De Pietri, R. and Derby, J. and De Rosa, R. and De Rossi, C. and DeSalvo, R. and de Varona, O. and Dhurandhar, S. and D\'{\i}az, M. C. and Dietrich, T. and Di Fiore, L. and Di Giovanni, M. and Di Girolamo, T. and Di Lieto, A. and Ding, B. and Di Pace, S. and Di Palma, I. and Di Renzo, F. and Dmitriev, A. and Doctor, Z. and Donovan, F. and Dooley, K. L. and Doravari, S. and Dorrington, I. and Downes, T. P. and Drago, M. and Driggers, J. C. and Du, Z. and Ducoin, J.-G. and Dupej, P. and Dwyer, S. E. and Easter, P. J. and Edo, T. B. and Edwards, M. C. and Effler, A. and Ehrens, P. and Eichholz, J. and Eikenberry, S. S. and Eisenmann, M. and Eisenstein, R. A. and Essick, R. C. and Estelles, H. and Estevez, D. and Etienne, Z. B. and Etzel, T. and Evans, M. and Evans, T. M. and Fafone, V. and Fair, H. and Fairhurst, S. and Fan, X. and Farinon, S. and Farr, B. and Farr, W. M. and Fauchon-Jones, E. J. and Favata, M. and Fays, M. and Fazio, M. and Fee, C. and Feicht, J. and Fejer, M. M. and Feng, F. and Fernandez-Galiana, A. and Ferrante, I. and Ferreira, E. C. and Ferreira, T. A. and Ferrini, F. and Fidecaro, F. and Fiori, I. and Fiorucci, D. and Fishbach, M. and Fisher, R. P. and Fishner, J. M. and Fitz-Axen, M. and Flaminio, R. and Fletcher, M. and Flynn, E. and Fong, H. and Font, J. A. and Forsyth, P. W. F. and Fournier, J.-D. and Frasca, S. and Frasconi, F. and Frei, Z. and Freise, A. and Frey, R. and Frey, V. and Fritschel, P. and Frolov, V. V. and Fulda, P. and Fyffe, M. and Gabbard, H. A. and Gadre, B. U. and Gaebel, S. M. and Gair, J. R. and Gammaitoni, L. and Ganija, M. R. and Gaonkar, S. G. and Garcia, A. and Garc\'{\i}a-Quir\'os, C. and Garufi, F. and Gateley, B. and Gaudio, S. and Gaur, G. and Gayathri, V. and Gemme, G. and Genin, E. and Gennai, A. and George, D. and George, J. and Gergely, L. and Germain, V. and Ghonge, S. and Ghosh, Abhirup and Ghosh, Archisman and Ghosh, S. and Giacomazzo, B. and Giaime, J. A. and Giardina, K. D. and Giazotto, A. and Gill, K. and Giordano, G. and Glover, L. and Godwin, P. and Goetz, E. and Goetz, R. and Goncharov, B. and Gonz\'alez, G. and Gonzalez Castro, J. M. and Gopakumar, A. and Gorodetsky, M. L. and Gossan, S. E. and Gosselin, M. and Gouaty, R. and Grado, A. and Graef, C. and Granata, M. and Grant, A. and Gras, S. and Grassia, P. and Gray, C. and Gray, R. and Greco, G. and Green, A. C. and Green, R. and Gretarsson, E. M. and Groot, P. and Grote, H. and Grunewald, S. and Gruning, P. and Guidi, G. M. and Gulati, H. K. and Guo, Y. and Gupta, A. and Gupta, M. K. and Gustafson, E. K. and Gustafson, R. and Haegel, L. and Halim, O. and Hall, B. R. and Hall, E. D. and Hamilton, E. Z. and Hammond, G. and Haney, M. and Hanke, M. M. and Hanks, J. and Hanna, C. and Hannam, M. D. and Hannuksela, O. A. and Hanson, J. and Hardwick, T. and Haris, K. and Harms, J. and Harry, G. M. and Harry, I. W. and Haster, C.-J. and Haughian, K. and Hayes, F. J. and Healy, J. and Heidmann, A. and Heintze, M. C. and Heitmann, H. and Hello, P. and Hemming, G. and Hendry, M. and Heng, I. S. and Hennig, J. and Heptonstall, A. W. and Hernandez Vivanco, Francisco and Heurs, M. and Hild, S. and Hinderer, T. and Hoak, D. and Hochheim, S. and Hofman, D. and Holgado, A. M. and Holland, N. A. and Holt, K. and Holz, D. E. and Hopkins, P. and Horst, C. and Hough, J. and Howell, E. J. and Hoy, C. G. and Hreibi, A. and Huang, Y. and Huerta, E. A. and Huet, D. and Hughey, B. and Hulko, M. and Husa, S. and Huttner, S. H. and Huynh-Dinh, T. and Idzkowski, B. and Iess, A. and Ingram, C. and Inta, R. and Intini, G. and Irwin, B. and Isa, H. N. and Isac, J.-M. and Isi, M. and Iyer, B. R. and Izumi, K. and Jacqmin, T. and Jadhav, S. J. and Jani, K. and Janthalur, N. N. and Jaranowski, P. and Jenkins, A. C. and Jiang, J. and Johnson, D. S. and Johnson-McDaniel, N. K. and Jones, A. W. and Jones, D. I. and Jones, R. and Jonker, R. J. G. and Ju, L. and Junker, J. and Kalaghatgi, C. V. and Kalogera, V. and Kamai, B. and Kandhasamy, S. and Kang, G. and Kanner, J. B. and Kapadia, S. J. and Karki, S. and Karvinen, K. S. and Kashyap, R. and Kasprzack, M. and Katsanevas, S. and Katsavounidis, E. and Katzman, W. and Kaufer, S. and Kawabe, K. and Keerthana, N. V. and K\'ef\'elian, F. and Keitel, D. and Kennedy, R. and Key, J. S. and Khalili, F. Y. and Khan, H. and Khan, I. and Khan, S. and Khan, Z. and Khazanov, E. A. and Khursheed, M. and Kijbunchoo, N. and Kim, Chunglee and Kim, J. C. and Kim, K. and Kim, W. and Kim, W. S. and Kim, Y.-M. and Kimball, C. and King, E. J. and King, P. J. and Kinley-Hanlon, M. and Kirchhoff, R. and Kissel, J. S. and Kleybolte, L. and Klika, J. H. and Klimenko, S. and Knowles, T. D. and Koch, P. and Koehlenbeck, S. M. and Koekoek, G. and Koley, S. and Kondrashov, V. and Kontos, A. and Koper, N. and Korobko, M. and Korth, W. Z. and Kowalska, I. and Kozak, D. B. and Kringel, V. and Krishnendu, N. and Kr\'olak, A. and Kuehn, G. and Kumar, A. and Kumar, P. and Kumar, R. and Kumar, S. and Kuo, L. and Kutynia, A. and Kwang, S. and Lackey, B. D. and Lai, K. H. and Lam, T. L. and Landry, M. and Lane, B. B. and Lang, R. N. and Lange, J. and Lantz, B. and Lanza, R. K. and Lartaux-Vollard, A. and Lasky, P. D. and Laxen, M. and Lazzarini, A. and Lazzaro, C. and Leaci, P. and Leavey, S. and Lecoeuche, Y. K. and Lee, C. H. and Lee, H. K. and Lee, H. M. and Lee, H. W. and Lee, J. and Lee, K. and Lehmann, J. and Lenon, A. and Leroy, N. and Letendre, N. and Levin, Y. and Li, J. and Li, K. J. L. and Li, T. G. F. and Li, X. and Lin, F. and Linde, F. and Linker, S. D. and Littenberg, T. B. and Liu, J. and Liu, X. and Lo, R. K. L. and Lockerbie, N. A. and London, L. T. and Longo, A. and Lorenzini, M. and Loriette, V. and Lormand, M. and Losurdo, G. and Lough, J. D. and Lousto, C. O. and Lovelace, G. and Lower, M. E. and L\"uck, H. and Lumaca, D. and Lundgren, A. P. and Lynch, R. and Ma, Y. and Macas, R. and Macfoy, S. and MacInnis, M. and Macleod, D. M. and Macquet, A. and Maga\~na-Sandoval, F. and Maga\~na Zertuche, L. and Magee, R. M. and Majorana, E. and Maksimovic, I. and Malik, A. and Man, N. and Mandic, V. and Mangano, V. and Mansell, G. L. and Manske, M. and Mantovani, M. and Marchesoni, F. and Marion, F. and M\'arka, S. and M\'arka, Z. and Markakis, C. and Markosyan, A. S. and Markowitz, A. and Maros, E. and Marquina, A. and Marsat, S. and Martelli, F. and Martin, I. W. and Martin, R. M. and Martynov, D. V. and Mason, K. and Massera, E. and Masserot, A. and Massinger, T. J. and Masso-Reid, M. and Mastrogiovanni, S. and Matas, A. and Matichard, F. and Matone, L. and Mavalvala, N. and Mazumder, N. and McCann, J. J. and McCarthy, R. and McClelland, D. E. and McCormick, S. and McCuller, L. and McGuire, S. C. and McIver, J. and McManus, D. J. and McRae, T. and McWilliams, S. T. and Meacher, D. and Meadors, G. D. and Mehmet, M. and Mehta, A. K. and Meidam, J. and Melatos, A. and Mendell, G. and Mercer, R. A. and Mereni, L. and Merilh, E. L. and Merzougui, M. and Meshkov, S. and Messenger, C. and Messick, C. and Metzdorff, R. and Meyers, P. M. and Miao, H. and Michel, C. and Middleton, H. and Mikhailov, E. E. and Milano, L. and Miller, A. L. and Miller, A. and Millhouse, M. and Mills, J. C. and Milovich-Goff, M. C. and Minazzoli, O. and Minenkov, Y. and Mishkin, A. and Mishra, C. and Mistry, T. and Mitra, S. and Mitrofanov, V. P. and Mitselmakher, G. and Mittleman, R. and Mo, G. and Moffa, D. and Mogushi, K. and Mohapatra, S. R. P. and Montani, M. and Moore, C. J. and Moraru, D. and Moreno, G. and Morisaki, S. and Mours, B. and Mow-Lowry, C. M. and Mukherjee, Arunava and Mukherjee, D. and Mukherjee, S. and Mukund, N. and Mullavey, A. and Munch, J. and Mu\~niz, E. A. and Muratore, M. and Murray, P. G. and Nagar, A. and Nardecchia, I. and Naticchioni, L. and Nayak, R. K. and Neilson, J. and Nelemans, G. and Nelson, T. J. N. and Nery, M. and Neunzert, A. and Ng, K. Y. and Ng, S. and Nguyen, P. and Nichols, D. and Nielsen, A. B. and Nissanke, S. and Nitz, A. and Nocera, F. and North, C. and Nuttall, L. K. and Obergaulinger, M. and Oberling, J. and O'Brien, B. D. and O'Dea, G. D. and Ogin, G. H. and Oh, J. J. and Oh, S. H. and Ohme, F. and Ohta, H. and Okada, M. A. and Oliver, M. and Oppermann, P. and Oram, Richard J. and O'Reilly, B. and Ormiston, R. G. and Ortega, L. F. and O'Shaughnessy, R. and Ossokine, S. and Ottaway, D. J. and Overmier, H. and Owen, B. J. and Pace, A. E. and Pagano, G. and Page, M. A. and Pai, A. and Pai, S. A. and Palamos, J. R. and Palashov, O. and Palomba, C. and Pal-Singh, A. and Pan, Huang-Wei and Pang, B. and Pang, P. T. H. and Pankow, C. and Pannarale, F. and Pant, B. C. and Paoletti, F. and Paoli, A. and Papa, M. A. and Parida, A. and Parker, W. and Pascucci, D. and Pasqualetti, A. and Passaquieti, R. and Passuello, D. and Patil, M. and Patricelli, B. and Pearlstone, B. L. and Pedersen, C. and Pedraza, M. and Pedurand, R. and Pele, A. and Penn, S. and Perego, A. and Perez, C. J. and Perreca, A. and Pfeiffer, H. P. and Phelps, M. and Phukon, K. S. and Piccinni, O. J. and Pichot, M. and Piergiovanni, F. and Pillant, G. and Pinard, L. and Pirello, M. and Pitkin, M. and Poggiani, R. and Pong, D. Y. T. and Ponrathnam, S. and Popolizio, P. and Porter, E. K. and Powell, J. and Prajapati, A. K. and Prasad, J. and Prasai, K. and Prasanna, R. and Pratten, G. and Prestegard, T. and Privitera, S. and Prodi, G. A. and Prokhorov, L. G. and Puncken, O. and Punturo, M. and Puppo, P. and P\"urrer, M. and Qi, H. and Quetschke, V. and Quinonez, P. J. and Quintero, E. A. and Quitzow-James, R. and Raab, F. J. and Radkins, H. and Radulescu, N. and Raffai, P. and Raja, S. and Rajan, C. and Rajbhandari, B. and Rakhmanov, M. and Ramirez, K. E. and Ramos-Buades, A. and Rana, Javed and Rao, K. and Rapagnani, P. and Raymond, V. and Razzano, M. and Read, J. and Regimbau, T. and Rei, L. and Reid, S. and Reitze, D. H. and Ren, W. and Ricci, F. and Richardson, C. J. and Richardson, J. W. and Ricker, P. M. and Riemenschneider, G. M. and Riles, K. and Rizzo, M. and Robertson, N. A. and Robie, R. and Robinet, F. and Rocchi, A. and Rolland, L. and Rollins, J. G. and Roma, V. J. and Romanelli, M. and Romano, R. and Romel, C. L. and Romie, J. H. and Rose, K. and Rosi\ifmmode \acute{n}\else \'{n}\fi{}ska, D. and Rosofsky, S. G. and Ross, M. P. and Rowan, S. and R\"udiger, A. and Ruggi, P. and Rutins, G. and Ryan, K. and Sachdev, S. and Sadecki, T. and Sakellariadou, M. and Salafia, O. and Salconi, L. and Saleem, M. and Salemi, F. and Samajdar, A. and Sammut, L. and Sanchez, E. J. and Sanchez, L. E. and Sanchis-Gual, N. and Sandberg, V. and Sanders, J. R. and Santiago, K. A. and Sarin, N. and Sassolas, B. and Sathyaprakash, B. S. and Saulson, P. R. and Sauter, O. and Savage, R. L. and Schale, P. and Scheel, M. and Scheuer, J. and Schmidt, P. and Schnabel, R. and Schofield, R. M. S. and Sch\"onbeck, A. and Schreiber, E. and Schulte, B. W. and Schutz, B. F. and Schwalbe, S. G. and Scott, J. and Scott, S. M. and Seidel, E. and Sellers, D. and Sengupta, A. S. and Sennett, N. and Sentenac, D. and Sequino, V. and Sergeev, A. and Setyawati, Y. and Shaddock, D. A. and Shaffer, T. and Shahriar, M. S. and Shaner, M. B. and Shao, L. and Sharma, P. and Shawhan, P. and Shen, H. and Shink, R. and Shoemaker, D. H. and Shoemaker, D. M. and ShyamSundar, S. and Siellez, K. and Sieniawska, M. and Sigg, D. and Silva, A. D. and Singer, L. P. and Singh, N. and Singhal, A. and Sintes, A. M. and Sitmukhambetov, S. and Skliris, V. and Slagmolen, B. J. J. and Slaven-Blair, T. J. and Smith, J. R. and Smith, R. J. E. and Somala, S. and Son, E. J. and Sorazu, B. and Sorrentino, F. and Souradeep, T. and Sowell, E. and Spencer, A. P. and Srivastava, A. K. and Srivastava, V. and Staats, K. and Stachie, C. and Standke, M. and Steer, D. A. and Steinke, M. and Steinlechner, J. and Steinlechner, S. and Steinmeyer, D. and Stevenson, S. P. and Stocks, D. and Stone, R. and Stops, D. J. and Strain, K. A. and Stratta, G. and Strigin, S. E. and Strunk, A. and Sturani, R. and Stuver, A. L. and Sudhir, V. and Summerscales, T. Z. and Sun, L. and Sunil, S. and Suresh, J. and Sutton, P. J. and Swinkels, B. L. and Szczepa\ifmmode \acute{n}\else \'{n}\fi{}czyk, M. J. and Tacca, M. and Tait, S. C. and Talbot, C. and Talukder, D. and Tanner, D. B. and T\'apai, M. and Taracchini, A. and Tasson, J. D. and Taylor, R. and Thies, F. and Thomas, M. and Thomas, P. and Thondapu, S. R. and Thorne, K. A. and Thrane, E. and Tiwari, Shubhanshu and Tiwari, Srishti and Tiwari, V. and Toland, K. and Tonelli, M. and Tornasi, Z. and Torres-Forn\'e, A. and Torrie, C. I. and T\"oyr\"a, D. and Travasso, F. and Traylor, G. and Tringali, M. C. and Trovato, A. and Trozzo, L. and Trudeau, R. and Tsang, K. W. and Tse, M. and Tso, R. and Tsukada, L. and Tsuna, D. and Tuyenbayev, D. and Ueno, K. and Ugolini, D. and Unnikrishnan, C. S. and Urban, A. L. and Usman, S. A. and Vahlbruch, H. and Vajente, G. and Valdes, G. and van Bakel, N. and van Beuzekom, M. and van den Brand, J. F. J. and Van Den Broeck, C. and Vander-Hyde, D. C. and van Heijningen, J. V. and van der Schaaf, L. and van Veggel, A. A. and Vardaro, M. and Varma, V. and Vass, S. and Vas\'uth, M. and Vecchio, A. and Vedovato, G. and Veitch, J. and Veitch, P. J. and Venkateswara, K. and Venugopalan, G. and Verkindt, D. and Vetrano, F. and Vicer\'e, A. and Viets, A. D. and Vine, D. J. and Vinet, J.-Y. and Vitale, S. and Vo, T. and Vocca, H. and Vorvick, C. and Vyatchanin, S. P. and Wade, A. R. and Wade, L. E. and Wade, M. and Walet, R. and Walker, M. and Wallace, L. and Walsh, S. and Wang, G. and Wang, H. and Wang, J. Z. and Wang, W. H. and Wang, Y. F. and Ward, R. L. and Warden, Z. A. and Warner, J. and Was, M. and Watchi, J. and Weaver, B. and Wei, L.-W. and Weinert, M. and Weinstein, A. J. and Weiss, R. and Wellmann, F. and Wen, L. and Wessel, E. K. and We\ss{}els, P. and Westhouse, J. W. and Wette, K. and Whelan, J. T. and White, L. V. and Whiting, B. F. and Whittle, C. and Wilken, D. M. and Williams, D. and Williamson, A. R. and Willis, J. L. and Willke, B. and Wimmer, M. H. and Winkler, W. and Wipf, C. C. and Wittel, H. and Woan, G. and Woehler, J. and Wofford, J. K. and Worden, J. and Wright, J. L. and Wu, D. S. and Wysocki, D. M. and Xiao, L. and Yamamoto, H. and Yancey, C. C. and Yang, L. and Yap, M. J. and Yazback, M. and Yeeles, D. W. and Yu, Hang and Yu, Haocun and Yuen, S. H. R. and Yvert, M. and Zadro\ifmmode \dot{z}\else \.{z}\fi{}ny, A. K. and Zanolin, M. and Zappa, F. and Zelenova, T. and Zendri, J.-P. and Zevin, M. and Zhang, J. and Zhang, L. and Zhang, T. and Zhao, C. and Zhou, M. and Zhou, Z. and Zhu, X. J. and Zimmerman, A. B. and Zlochower, Y. and Zucker, M. E. and Zweizig, J.},
  collaboration = {LIGO Scientific Collaboration and Virgo Collaboration},
  journal = {Phys. Rev. X},
  volume = {9},
  issue = {3},
  pages = {031040},
  numpages = {49},
  year = {2019},
  month = {Sep},
  publisher = {American Physical Society},
  doi = {10.1103/PhysRevX.9.031040},
  url = {https://link.aps.org/doi/10.1103/PhysRevX.9.031040}
}

@article{LIGO_O4,
  title = "{Advanced LIGO detector performance in the fourth observing run}",
  author = {Capote, E. and Jia, W. and Aritomi, N. and Nakano, M. and Xu, V. and Abbott, R. and Abouelfettouh, I. and Adhikari, R. X. and Ananyeva, A. and Appert, S. and Apple, S. K. and Arai, K. and Aston, S. M. and Ball, M. and Ballmer, S. W. and Barker, D. and Barsotti, L. and Berger, B. K. and Betzwieser, J. and Bhattacharjee, D. and Billingsley, G. and Biscans, S. and Blair, C. D. and Bode, N. and Bonilla, E. and Bossilkov, V. and Branch, A. and Brooks, A. F. and Brown, D. D. and Bryant, J. and Cahillane, C. and Cao, H. and Clara, F. and Collins, J. and Compton, C. M. and Cottingham, R. and Coyne, D. C. and Crouch, R. and Csizmazia, J. and Cumming, A. and Dartez, L. P. and Davis, D. and Demos, N. and Dohmen, E. and Driggers, J. C. and Dwyer, S. E. and Effler, A. and Ejlli, A. and Etzel, T. and Evans, M. and Feicht, J. and Frey, R. and Frischhertz, W. and Fritschel, P. and Frolov, V. V. and Fuentes-Garcia, M. and Fulda, P. and Fyffe, M. and Ganapathy, D. and Gateley, B. and Gayer, T. and Giaime, J. A. and Giardina, K. D. and Glanzer, J. and Goetz, E. and Goetz, R. and Goodwin-Jones, A. W. and Gras, S. and Gray, C. and Griffith, D. and Grote, H. and Guidry, T. and Gurs, J. and Hall, E. D. and Hanks, J. and Hanson, J. and Heintze, M. C. and Helmling-Cornell, A. F. and Holland, N. A. and Hoyland, D. and Huang, H. Y. and Inoue, Y. and James, A. L. and Jamies, A. and Jennings, A. and Jones, D. H. and Kabagoz, H. B. and Karat, S. and Karki, S. and Kasprzack, M. and Kawabe, K. and Kijbunchoo, N. and King, P. J. and Kissel, J. S. and Komori, K. and Kontos, A. and Kumar, Rahul and Kuns, K. and Landry, M. and Lantz, B. and Laxen, M. and Lee, K. and Lesovsky, M. and Villarreal, F. Llamas and Lormand, M. and Loughlin, H. A. and Macas, R. and MacInnis, M. and Makarem, C. N. and Mannix, B. and Mansell, G. L. and Martin, R. M. and Mason, K. and Matichard, F. and Mavalvala, N. and Maxwell, N. and McCarrol, G. and McCarthy, R. and McClelland, D. E. and McCormick, S. and McRae, T. and Mera, F. and Merilh, E. L. and Meylahn, F. and Mittleman, R. and Moraru, D. and Moreno, G. and Mullavey, A. and Nelson, T. J. N. and Neunzert, A. and Notte, J. and Oberling, J. and O'Hanlon, T. and Osthelder, C. and Ottaway, D. J. and Overmier, H. and Parker, W. and Patane, O. and Pele, A. and Pham, H. and Pirello, M. and Pullin, J. and Quetschke, V. and Ramirez, K. E. and Ransom, K. and Reyes, J. and Richardson, J. W. and Robinson, M. and Rollins, J. G. and Romel, C. L. and Romie, J. H. and Ross, M. P. and Ryan, K. and Sadecki, T. and Sanchez, A. and Sanchez, E. J. and Sanchez, L. E. and Savage, R. L. and Schaetzl, D. and Schiworski, M. G. and Schnabel, R. and Schofield, R. M. S. and Schwartz, E. and Sellers, D. and Shaffer, T. and Short, R. W. and Sigg, D. and Slagmolen, B. J. J. and Soike, C. and Soni, S. and Srivastava, V. and Sun, L. and Tanner, D. B. and Thomas, M. and Thomas, P. and Thorne, K. A. and Todd, M. R. and Torrie, C. I. and Traylor, G. and Ubhi, A. S. and Vajente, G. and Vanosky, J. and Vecchio, A. and Veitch, P. J. and Vibhute, A. M. and von Reis, E. R. G. and Warner, J. and Weaver, B. and Weiss, R. and Whittle, C. and Willke, B. and Wipf, C. C. and Wright, J. L. and Yamamoto, H. and Zhang, L. and Zucker, M. E.},
  journal = {Phys. Rev. D},
  volume = {111},
  issue = {6},
  pages = {062002},
  numpages = {30},
  year = {2025},
  month = {Mar},
  publisher = {American Physical Society},
  doi = {10.1103/PhysRevD.111.062002},
  url = {https://link.aps.org/doi/10.1103/PhysRevD.111.062002}
}

@article{GWTC2,
  title = "{GWTC-2.1: Deep extended catalog of compact binary coalescences observed by LIGO and Virgo during the first half of the third observing run}",
  author = {Abbott, R. and Abbott, T. D. and Acernese, F. and Ackley, K. and Adams, C. and Adhikari, N. and Adhikari, R. X. and Adya, V. B. and Affeldt, C. and Agarwal, D. and Agathos, M. and Agatsuma, K. and Aggarwal, N. and Aguiar, O. D. and Aiello, L. and Ain, A. and Ajith, P. and Albanesi, S. and Allocca, A. and Altin, P. A. and Amato, A. and Anand, C. and Anand, S. and Ananyeva, A. and Anderson, S. B. and Anderson, W. G. and Andrade, T. and Andres, N. and Andri\ifmmode \acute{c}\else \'{c}\fi{}, T. and Angelova, S. V. and Ansoldi, S. and Antelis, J. M. and Antier, S. and Appert, S. and Arai, K. and Araya, M. C. and Areeda, J. S. and Ar\`ene, M. and Arnaud, N. and Aronson, S. M. and Arun, K. G. and Asali, Y. and Ashton, G. and Assiduo, M. and Aston, S. M. and Astone, P. and Aubin, F. and Austin, C. and Babak, S. and Badaracco, F. and Bader, M. K. M. and Badger, C. and Bae, S. and Baer, A. M. and Bagnasco, S. and Bai, Y. and Baird, J. and Ball, M. and Ballardin, G. and Ballmer, S. W. and Balsamo, A. and Baltus, G. and Banagiri, S. and Bankar, D. and Barayoga, J. C. and Barbieri, C. and Barish, B. C. and Barker, D. and Barneo, P. and Barone, F. and Barr, B. and Barsotti, L. and Barsuglia, M. and Barta, D. and Bartlett, J. and Barton, M. A. and Bartos, I. and Bassiri, R. and Basti, A. and Bawaj, M. and Bayley, J. C. and Baylor, A. C. and Bazzan, M. and B\'ecsy, B. and Bedakihale, V. M. and Bejger, M. and Belahcene, I. and Benedetto, V. and Beniwal, D. and Bennett, T. F. and Bentley, J. D. and BenYaala, M. and Bergamin, F. and Berger, B. K. and Bernuzzi, S. and Berry, C. P. L. and Bersanetti, D. and Bertolini, A. and Betzwieser, J. and Beveridge, D. and Bhandare, R. and Bhardwaj, U. and Bhattacharjee, D. and Bhaumik, S. and Bilenko, I. A. and Billingsley, G. and Bini, S. and Birney, I. A. and Birnholtz, O. and Biscans, S. and Bischi, M. and Biscoveanu, S. and Bisht, A. and Biswas, B. and Bitossi, M. and Bizouard, M.-A. and Blackburn, J. K. and Blair, C. D. and Blair, D. G. and Blair, R. M. and Bobba, F. and Bode, N. and Boer, M. and Bogaert, G. and Boldrini, M. and Bonavena, L. D. and Bondu, F. and Bonilla, E. and Bonnand, R. and Booker, P. and Boom, B. A. and Bork, R. and Boschi, V. and Bose, N. and Bose, S. and Bossilkov, V. and Boudart, V. and Bouffanais, Y. and Bozzi, A. and Bradaschia, C. and Brady, P. R. and Bramley, A. and Branch, A. and Branchesi, M. and Brau, J. E. and Breschi, M. and Briant, T. and Briggs, J. H. and Brillet, A. and Brinkmann, M. and Brockill, P. and Brooks, A. F. and Brooks, J. and Brown, D. D. and Brunett, S. and Bruno, G. and Bruntz, R. and Bryant, J. and Bulik, T. and Bulten, H. J. and Buonanno, A. and Buscicchio, R. and Buskulic, D. and Buy, C. and Byer, R. L. and Cadonati, L. and Cagnoli, G. and Cahillane, C. and Bustillo, J. Calder\'on and Callaghan, J. D. and Callister, T. A. and Calloni, E. and Cameron, J. and Camp, J. B. and Canepa, M. and Canevarolo, S. and Cannavacciuolo, M. and Cannon, K. C. and Cao, H. and Capote, E. and Carapella, G. and Carbognani, F. and Carlin, J. B. and Carney, M. F. and Carpinelli, M. and Carrillo, G. and Carullo, G. and Carver, T. L. and Diaz, J. Casanueva and Casentini, C. and Castaldi, G. and Caudill, S. and Cavagli\`a, M. and Cavalier, F. and Cavalieri, R. and Ceasar, M. and Cella, G. and Cerd\'a-Dur\'an, P. and Cesarini, E. and Chaibi, W. and Chakravarti, K. and Subrahmanya, S. Chalathadka and Champion, E. and Chan, C.-H. and Chan, C. and Chan, C. L. and Chan, K. and Chandra, K. and Chanial, P. and Chao, S. and Charlton, P. and Chase, E. A. and Chassande-Mottin, E. and Chatterjee, C. and Chatterjee, Debarati and Chatterjee, Deep and Chattopadhyay, D. and Chaturvedi, M. and Chaty, S. and Chatziioannou, K. and Chen, H. Y. and Chen, J. and Chen, X. and Chen, Y. and Chen, Z. and Cheng, H. and Cheong, C. K. and Cheung, H. Y. and Chia, H. Y. and Chiadini, F. and Chiarini, G. and Chierici, R. and Chincarini, A. and Chiofalo, M. L. and Chiummo, A. and Cho, G. and Cho, H. S. and Choudhary, R. K. and Choudhary, S. and Christensen, N. and Chu, Q. and Chua, S. and Chung, K. W. and Ciani, G. and Ciecielag, P. and Cie\ifmmode \acute{s}\else \'{s}\fi{}lar, M. and Cifaldi, M. and Ciobanu, A. A. and Ciolfi, R. and Cipriano, F. and Cirone, A. and Clara, F. and Clark, E. N. and Clark, J. A. and Clarke, L. and Clearwater, P. and Clesse, S. and Cleva, F. and Coccia, E. and Codazzo, E. and Cohadon, P.-F. and Cohen, D. E. and Cohen, L. and Colleoni, M. and Collette, C. G. and Colombo, A. and Colpi, M. and Compton, C. M. and Constancio, M. and Conti, L. and Cooper, S. J. and Corban, P. and Corbitt, T. R. and Cordero-Carri\'on, I. and Corezzi, S. and Corley, K. R. and Cornish, N. and Corre, D. and Corsi, A. and Cortese, S. and Costa, C. A. and Cotesta, R. and Coughlin, M. W. and Coulon, J.-P. and Countryman, S. T. and Cousins, B. and Couvares, P. and Coward, D. M. and Cowart, M. J. and Coyne, D. C. and Coyne, R. and Creighton, J. D. E. and Creighton, T. D. and Criswell, A. W. and Croquette, M. and Crowder, S. G. and Cudell, J. R. and Cullen, T. J. and Cumming, A. and Cummings, R. and Cunningham, L. and Cuoco, E. and Cury\l{}o, M. and Dabadie, P. and Canton, T. Dal and Dall'Osso, S. and D\'alya, G. and Dana, A. and DaneshgaranBajastani, L. M. and D'Angelo, B. and Danila, B. and Danilishin, S. and D'Antonio, S. and Danzmann, K. and Darsow-Fromm, C. and Dasgupta, A. and Datrier, L. E. H. and Datta, S. and Dattilo, V. and Dave, I. and Davier, M. and Davies, G. S. and Davis, D. and Davis, M. C. and Daw, E. J. and Dean, R. and DeBra, D. and Deenadayalan, M. and Degallaix, J. and De Laurentis, M. and Del\'eglise, S. and Del Favero, V. and De Lillo, F. and De Lillo, N. and Del Pozzo, W. and DeMarchi, L. M. and De Matteis, F. and D'Emilio, V. and Demos, N. and Dent, T. and Depasse, A. and De Pietri, R. and De Rosa, R. and De Rossi, C. and DeSalvo, R. and De Simone, R. and Dhurandhar, S. and D\'{\i}az, M. C. and Diaz-Ortiz, M. and Didio, N. A. and Dietrich, T. and Di Fiore, L. and Di Fronzo, C. and Di Giorgio, C. and Di Giovanni, F. and Di Giovanni, M. and Di Girolamo, T. and Di Lieto, A. and Ding, B. and Di Pace, S. and Di Palma, I. and Di Renzo, F. and Divakarla, A. K. and Divyajyoti and Dmitriev, A. and Doctor, Z. and D'Onofrio, L. and Donovan, F. and Dooley, K. L. and Doravari, S. and Dorrington, I. and Drago, M. and Driggers, J. C. and Drori, Y. and Ducoin, J.-G. and Dupej, P. and Durante, O. and D'Urso, D. and Duverne, P.-A. and Dwyer, S. E. and Eassa, C. and Easter, P. J. and Ebersold, M. and Eckhardt, T. and Eddolls, G. and Edelman, B. and Edo, T. B. and Edy, O. and Effler, A. and Eichholz, J. and Eikenberry, S. S. and Eisenmann, M. and Eisenstein, R. A. and Ejlli, A. and Engelby, E. and Errico, L. and Essick, R. C. and Estell\'es, H. and Estevez, D. and Etienne, Z. and Etzel, T. and Evans, M. and Evans, T. M. and Ewing, B. E. and Fafone, V. and Fair, H. and Fairhurst, S. and Fanning, S. P. and Farah, A. M. and Farinon, S. and Farr, B. and Farr, W. M. and Farrow, N. W. and Fauchon-Jones, E. J. and Favaro, G. and Favata, M. and Fays, M. and Fazio, M. and Feicht, J. and Fejer, M. M. and Fenyvesi, E. and Ferguson, D. L. and Fernandez-Galiana, A. and Ferrante, I. and Ferreira, T. A. and Fidecaro, F. and Figura, P. and Fiori, I. and Fishbach, M. and Fisher, R. P. and Fittipaldi, R. and Fiumara, V. and Flaminio, R. and Floden, E. and Fong, H. and Font, J. A. and Fornal, B. and Forsyth, P. W. F. and Franke, A. and Frasca, S. and Frasconi, F. and Frederick, C. and Freed, J. P. and Frei, Z. and Freise, A. and Frey, R. and Fritschel, P. and Frolov, V. V. and Fronz\'e, G. G. and Fulda, P. and Fyffe, M. and Gabbard, H. A. and Gabella, W. and Gadre, B. U. and Gair, J. R. and Gais, J. and Galaudage, S. and Gamba, R. and Ganapathy, D. and Ganguly, A. and Gaonkar, S. G. and Garaventa, B. and Garc\'{\i}a, F. and Garc\'{\i}a-N\'u\~nez, C. and Garc\'{\i}a-Quir\'os, C. and Garufi, F. and Gateley, B. and Gaudio, S. and Gayathri, V. and Gemme, G. and Gennai, A. and George, J. and George, R. N. and Gerberding, O. and Gergely, L. and Gewecke, P. and Ghonge, S. and Ghosh, Abhirup and Ghosh, Archisman and Ghosh, Shaon and Ghosh, Shrobana and Giacomazzo, B. and Giacoppo, L. and Giaime, J. A. and Giardina, K. D. and Gibson, D. R. and Gier, C. and Giesler, M. and Giri, P. and Gissi, F. and Glanzer, J. and Gleckl, A. E. and Godwin, P. and Goetz, E. and Goetz, R. and Gohlke, N. and Goncharov, B. and Gonz\'alez, G. and Gopakumar, A. and Gosselin, M. and Gouaty, R. and Gould, D. W. and Grace, B. and Grado, A. and Granata, M. and Granata, V. and Grant, A. and Gras, S. and Grassia, P. and Gray, C. and Gray, R. and Greco, G. and Green, A. C. and Green, R. and Gretarsson, A. M. and Gretarsson, E. M. and Griffith, D. and Griffiths, W. and Griggs, H. L. and Grignani, G. and Grimaldi, A. and Grimm, S. J. and Grote, H. and Grunewald, S. and Gruning, P. and Guerra, D. and Guidi, G. M. and Guimaraes, A. R. and Guix\'e, G. and Gulati, H. K. and Guo, H.-K. and Guo, Y. and Gupta, Anchal and Gupta, Anuradha and Gupta, P. and Gustafson, E. K. and Gustafson, R. and Guzman, F. and Haegel, L. and Halim, O. and Hall, E. D. and Hamilton, E. Z. and Hammond, G. and Haney, M. and Hanks, J. and Hanna, C. and Hannam, M. D. and Hannuksela, O. and Hansen, H. and Hansen, T. J. and Hanson, J. and Harder, T. and Hardwick, T. and Haris, K. and Harms, J. and Harry, G. M. and Harry, I. W. and Hartwig, D. and Haskell, B. and Hasskew, R. K. and Haster, C.-J. and Haughian, K. and Hayes, F. J. and Healy, J. and Heidmann, A. and Heidt, A. and Heintze, M. C. and Heinze, J. and Heinzel, J. and Heitmann, H. and Hellman, F. and Hello, P. and Helmling-Cornell, A. F. and Hemming, G. and Hendry, M. and Heng, I. S. and Hennes, E. and Hennig, J. and Hennig, M. H. and Hernandez, A. G. and Vivanco, F. Hernandez and Heurs, M. and Hild, S. and Hill, P. and Hines, A. S. and Hochheim, S. and Hofman, D. and Hohmann, J. N. and Holcomb, D. G. and Holland, N. A. and Holley-Bockelmann, K. and Hollows, I. J. and Holmes, Z. J. and Holt, K. and Holz, D. E. and Hopkins, P. and Hough, J. and Hourihane, S. and Howell, E. J. and Hoy, C. G. and Hoyland, D. and Hreibi, A. and Hsu, Y. and Huang, Y. and H\"ubner, M. T. and Huddart, A. D. and Hughey, B. and Hui, V. and Husa, S. and Huttner, S. H. and Huxford, R. and Huynh-Dinh, T. and Idzkowski, B. and Iess, A. and Ingram, C. and Isi, M. and Isleif, K. and Iyer, B. R. and JaberianHamedan, V. and Jacqmin, T. and Jadhav, S. J. and Jadhav, S. P. and James, A. L. and Jan, A. Z. and Jani, K. and Janquart, J. and Janssens, K. and Janthalur, N. N. and Jaranowski, P. and Jariwala, D. and Jaume, R. and Jenkins, A. C. and Jenner, K. and Jeunon, M. and Jia, W. and Johns, G. R. and Johnson-McDaniel, N. K. and Jones, A. W. and Jones, D. I. and Jones, J. D. and Jones, P. and Jones, R. and Jonker, R. J. G. and Ju, L. and Junker, J. and Juste, V. and Kalaghatgi, C. V. and Kalogera, V. and Kamai, B. and Kandhasamy, S. and Kang, G. and Kanner, J. B. and Kao, Y. and Kapadia, S. J. and Kapasi, D. P. and Karat, S. and Karathanasis, C. and Karki, S. and Kashyap, R. and Kasprzack, M. and Kastaun, W. and Katsanevas, S. and Katsavounidis, E. and Katzman, W. and Kaur, T. and Kawabe, K. and K\'ef\'elian, F. and Keitel, D. and Key, J. S. and Khadka, S. and Khalili, F. Y. and Khan, S. and Khazanov, E. A. and Khetan, N. and Khursheed, M. and Kijbunchoo, N. and Kim, C. and Kim, J. C. and Kim, K. and Kim, W. S. and Kim, Y.-M. and Kimball, C. and Kinley-Hanlon, M. and Kirchhoff, R. and Kissel, J. S. and Kleybolte, L. and Klimenko, S. and Knee, A. M. and Knowles, T. D. and Knyazev, E. and Koch, P. and Koekoek, G. and Koley, S. and Kolitsidou, P. and Kolstein, M. and Komori, K. and Kondrashov, V. and Kontos, A. and Koper, N. and Korobko, M. and Kovalam, M. and Kozak, D. B. and Kringel, V. and Krishnendu, N. V. and Kr\'olak, A. and Kuehn, G. and Kuei, F. and Kuijer, P. and Kulkarni, S. and Kumar, A. and Kumar, P. and Kumar, Rahul and Kumar, Rakesh and Kuns, K. and Kuwahara, S. and Lagabbe, P. and Laghi, D. and Lalande, E. and Lam, T. L. and Lamberts, A. and Landry, M. and Lane, B. B. and Lang, R. N. and Lange, J. and Lantz, B. and La Rosa, I. and Lartaux-Vollard, A. and Lasky, P. D. and Laxen, M. and Lazzarini, A. and Lazzaro, C. and Leaci, P. and Leavey, S. and Lecoeuche, Y. K. and Lee, H. M. and Lee, H. W. and Lee, J. and Lee, K. and Lehmann, J. and Lema\^{\i}tre, A. and Leroy, N. and Letendre, N. and Levesque, C. and Levin, Y. and Leviton, J. N. and Leyde, K. and Li, A. K. Y. and Li, B. and Li, J. and Li, T. G. F. and Li, X. and Linde, F. and Linker, S. D. and Linley, J. N. and Littenberg, T. B. and Liu, J. and Liu, K. and Liu, X. and Llamas, F. and Llorens-Monteagudo, M. and Lo, R. K. L. and Lockwood, A. and London, L. T. and Longo, A. and Lopez, D. and Portilla, M. Lopez and Lorenzini, M. and Loriette, V. and Lormand, M. and Losurdo, G. and Lott, T. P. and Lough, J. D. and Lousto, C. O. and Lovelace, G. and Lucaccioni, J. F. and L\"uck, H. and Lumaca, D. and Lundgren, A. P. and Lynam, J. E. and Macas, R. and MacInnis, M. and Macleod, D. M. and MacMillan, I. A. O. and Macquet, A. and Hernandez, I. Maga\~na and Magazz\`u, C. and Magee, R. M. and Maggiore, R. and Magnozzi, M. and Mahesh, S. and Majorana, E. and Makarem, C. and Maksimovic, I. and Maliakal, S. and Malik, A. and Man, N. and Mandic, V. and Mangano, V. and Mango, J. L. and Mansell, G. L. and Manske, M. and Mantovani, M. and Mapelli, M. and Marchesoni, F. and Marion, F. and Mark, Z. and M\'arka, S. and M\'arka, Z. and Markakis, C. and Markosyan, A. S. and Markowitz, A. and Maros, E. and Marquina, A. and Marsat, S. and Martelli, F. and Martin, I. W. and Martin, R. M. and Martinez, M. and Martinez, V. A. and Martinez, V. and Martinovic, K. and Martynov, D. V. and Marx, E. J. and Masalehdan, H. and Mason, K. and Massera, E. and Masserot, A. and Massinger, T. J. and Masso-Reid, M. and Mastrogiovanni, S. and Matas, A. and Mateu-Lucena, M. and Matichard, F. and Matiushechkina, M. and Mavalvala, N. and McCann, J. J. and McCarthy, R. and McClelland, D. E. and McClincy, P. K. and McCormick, S. and McCuller, L. and McGhee, G. I. and McGuire, S. C. and McIsaac, C. and McIver, J. and McRae, T. and McWilliams, S. T. and Meacher, D. and Mehmet, M. and Mehta, A. K. and Meijer, Q. and Melatos, A. and Melchor, D. A. and Mendell, G. and Menendez-Vazquez, A. and Menoni, C. S. and Mercer, R. A. and Mereni, L. and Merfeld, K. and Merilh, E. L. and Merritt, J. D. and Merzougui, M. and Meshkov, S. and Messenger, C. and Messick, C. and Meyers, P. M. and Meylahn, F. and Mhaske, A. and Miani, A. and Miao, H. and Michaloliakos, I. and Michel, C. and Middleton, H. and Milano, L. and Miller, A. and Miller, A. L. and Miller, B. and Millhouse, M. and Mills, J. C. and Milotti, E. and Minazzoli, O. and Minenkov, Y. and Mir, Ll. M. and Miravet-Ten\'es, M. and Mishra, C. and Mishra, T. and Mistry, T. and Mitra, S. and Mitrofanov, V. P. and Mitselmakher, G. and Mittleman, R. and Mo, Geoffrey and Moguel, E. and Mogushi, K. and Mohapatra, S. R. P. and Mohite, S. R. and Molina, I. and Molina-Ruiz, M. and Mondin, M. and Montani, M. and Moore, C. J. and Moraru, D. and Morawski, F. and More, A. and Moreno, C. and Moreno, G. and Morisaki, S. and Mours, B. and Mow-Lowry, C. M. and Mozzon, S. and Muciaccia, F. and Mukherjee, Arunava and Mukherjee, D. and Mukherjee, Soma and Mukherjee, Subroto and Mukherjee, Suvodip and Mukund, N. and Mullavey, A. and Munch, J. and Mu\~niz, E. A. and Murray, P. G. and Musenich, R. and Muusse, S. and Nadji, S. L. and Nagar, A. and Napolano, V. and Nardecchia, I. and Naticchioni, L. and Nayak, B. and Nayak, R. K. and Neil, B. F. and Neilson, J. and Nelemans, G. and Nelson, T. J. N. and Nery, M. and Neubauer, P. and Neunzert, A. and Ng, K. Y. and Ng, S. W. S. and Nguyen, C. and Nguyen, P. and Nguyen, T. and Nichols, S. A. and Nissanke, S. and Nitoglia, E. and Nocera, F. and Norman, M. and North, C. and Nuttall, L. K. and Oberling, J. and O'Brien, B. D. and O'Dell, J. and Oelker, E. and Oganesyan, G. and Oh, J. J. and Oh, S. H. and Ohme, F. and Ohta, H. and Okada, M. A. and Olivetto, C. and Oram, R. and O'Reilly, B. and Ormiston, R. G. and Ormsby, N. D. and Ortega, L. F. and O'Shaughnessy, R. and O'Shea, E. and Ossokine, S. and Osthelder, C. and Ottaway, D. J. and Overmier, H. and Pace, A. E. and Pagano, G. and Page, M. A. and Pagliaroli, G. and Pai, A. and Pai, S. A. and Palamos, J. R. and Palashov, O. and Palomba, C. and Pan, H. and Panda, P. K. and Pang, P. T. H. and Pankow, C. and Pannarale, F. and Pant, B. C. and Panther, F. H. and Paoletti, F. and Paoli, A. and Paolone, A. and Park, H. and Parker, W. and Pascucci, D. and Pasqualetti, A. and Passaquieti, R. and Passuello, D. and Patel, M. and Pathak, M. and Patricelli, B. and Patron, A. S. and Patrone, S. and Paul, S. and Payne, E. and Pedraza, M. and Pegoraro, M. and Pele, A. and Penn, S. and Perego, A. and Pereira, A. and Pereira, T. and Perez, C. J. and P\'erigois, C. and Perkins, C. C. and Perreca, A. and Perri\`es, S. and Petermann, J. and Petterson, D. and Pfeiffer, H. P. and Pham, K. A. and Phukon, K. S. and Piccinni, O. J. and Pichot, M. and Piendibene, M. and Piergiovanni, F. and Pierini, L. and Pierro, V. and Pillant, G. and Pillas, M. and Pilo, F. and Pinard, L. and Pinto, I. M. and Pinto, M. and Piotrzkowski, K. and Pirello, M. and Pitkin, M. D. and Placidi, E. and Planas, L. and Plastino, W. and Pluchar, C. and Poggiani, R. and Polini, E. and Pong, D. Y. T. and Ponrathnam, S. and Popolizio, P. and Porter, E. K. and Poulton, R. and Powell, J. and Pracchia, M. and Pradier, T. and Prajapati, A. K. and Prasai, K. and Prasanna, R. and Pratten, G. and Principe, M. and Prodi, G. A. and Prokhorov, L. and Prosposito, P. and Prudenzi, L. and Puecher, A. and Punturo, M. and Puosi, F. and Puppo, P. and P\"urrer, M. and Qi, H. and Quetschke, V. and Quitzow-James, R. and Raab, F. J. and Raaijmakers, G. and Radkins, H. and Radulesco, N. and Raffai, P. and Rail, S. X. and Raja, S. and Rajan, C. and Ramirez, K. E. and Ramirez, T. D. and Ramos-Buades, A. and Rana, J. and Rapagnani, P. and Rapol, U. D. and Ray, A. and Raymond, V. and Raza, N. and Razzano, M. and Read, J. and Rees, L. A. and Regimbau, T. and Rei, L. and Reid, S. and Reid, S. W. and Reitze, D. H. and Relton, P. and Renzini, A. and Rettegno, P. and Reza, A. and Rezac, M. and Ricci, F. and Richards, D. and Richardson, J. W. and Richardson, L. and Riemenschneider, G. and Riles, K. and Rinaldi, S. and Rink, K. and Rizzo, M. and Robertson, N. A. and Robie, R. and Robinet, F. and Rocchi, A. and Rodriguez, S. and Rolland, L. and Rollins, J. G. and Romanelli, M. and Romano, R. and Romel, C. L. and Romero-Rodr\'{\i}guez, A. and Romero-Shaw, I. M. and Romie, J. H. and Ronchini, S. and Rosa, L. and Rose, C. A. and Rosell, M. J. B. and Rosi\ifmmode \acute{n}\else \'{n}\fi{}ska, D. and Ross, M. P. and Rowan, S. and Rowlinson, S. J. and Roy, S. and Roy, Santosh and Roy, Soumen and Rozza, D. and Ruggi, P. and Ruiz-Rocha, K. and Ryan, K. and Sachdev, S. and Sadecki, T. and Sadiq, J. and Sakellariadou, M. and Salafia, O. S. and Salconi, L. and Saleem, M. and Salemi, F. and Samajdar, A. and Sanchez, E. J. and Sanchez, J. H. and Sanchez, L. E. and Sanchis-Gual, N. and Sanders, J. R. and Sanuy, A. and Saravanan, T. R. and Sarin, N. and Sassolas, B. and Satari, H. and Sauter, O. and Savage, R. L. and Sawant, D. and Sawant, H. L. and Sayah, S. and Schaetzl, D. and Scheel, M. and Scheuer, J. and Schiworski, M. and Schmidt, P. and Schmidt, S. and Schnabel, R. and Schneewind, M. and Schofield, R. M. S. and Sch\"onbeck, A. and Schulte, B. W. and Schutz, B. F. and Schwartz, E. and Scott, J. and Scott, S. M. and Seglar-Arroyo, M. and Sellers, D. and Sengupta, A. S. and Sentenac, D. and Seo, E. G. and Sequino, V. and Sergeev, A. and Setyawati, Y. and Shaffer, T. and Shahriar, M. S. and Shams, B. and Sharma, A. and Sharma, P. and Shawhan, P. and Shcheblanov, N. S. and Shikauchi, M. and Shoemaker, D. H. and Shoemaker, D. M. and ShyamSundar, S. and Sieniawska, M. and Sigg, D. and Singer, L. P. and Singh, D. and Singh, N. and Singha, A. and Sintes, A. M. and Sipala, V. and Skliris, V. and Slagmolen, B. J. J. and Slaven-Blair, T. J. and Smetana, J. and Smith, J. R. and Smith, R. J. E. and Soldateschi, J. and Somala, S. N. and Son, E. J. and Soni, K. and Soni, S. and Sordini, V. and Sorrentino, F. and Sorrentino, N. and Soulard, R. and Souradeep, T. and Sowell, E. and Spagnuolo, V. and Spencer, A. P. and Spera, M. and Srinivasan, R. and Srivastava, A. K. and Srivastava, V. and Staats, K. and Stachie, C. and Steer, D. A. and Steinhoff, J. and Steinlechner, J. and Steinlechner, S. and Stevenson, S. and Stops, D. J. and Stover, M. and Strain, K. A. and Strang, L. C. and Stratta, G. and Strunk, A. and Sturani, R. and Stuver, A. L. and Sudhagar, S. and Sudhir, V. and Suh, H. G. and Summerscales, T. Z. and Sun, H. and Sun, L. and Sunil, S. and Sur, A. and Suresh, J. and Sutton, P. J. and Swinkels, B. L. and Szczepa\ifmmode \acute{n}\else \'{n}\fi{}czyk, M. J. and Szewczyk, P. and Tacca, M. and Tait, S. C. and Talbot, C. J. and Talbot, C. and Tanasijczuk, A. J. and Tanner, D. B. and Tao, D. and Tao, L. and Mart\'{\i}n, E. N. Tapia San and Taranto, C. and Tasson, J. D. and Tenorio, R. and Terhune, J. E. and Terkowski, L. and Thirugnanasambandam, M. P. and Thomas, L. and Thomas, M. and Thomas, P. and Thompson, J. E. and Thondapu, S. R. and Thorne, K. A. and Thrane, E. and Tiwari, Shubhanshu and Tiwari, Srishti and Tiwari, V. and Toivonen, A. M. and Toland, K. and Tolley, A. E. and Tonelli, M. and Torres-Forn\'e, A. and Torrie, C. I. and e Melo, I. Tosta and T\"oyr\"a, D. and Trapananti, A. and Travasso, F. and Traylor, G. and Trevor, M. and Tringali, M. C. and Tripathee, A. and Troiano, L. and Trovato, A. and Trozzo, L. and Trudeau, R. J. and Tsai, D. S. and Tsai, D. and Tsang, K. W. and Tse, M. and Tso, R. and Tsukada, L. and Tsuna, D. and Tsutsui, T. and Turbang, K. and Turconi, M. and Ubhi, A. S. and Udall, R. P. and Ueno, K. and Unnikrishnan, C. S. and Urban, A. L. and Utina, A. and Vahlbruch, H. and Vajente, G. and Vajpeyi, A. and Valdes, G. and Valentini, M. and Valsan, V. and van Bakel, N. and van Beuzekom, M. and van den Brand, J. F. J. and Van Den Broeck, C. and Vander-Hyde, D. C. and van der Schaaf, L. and van Heijningen, J. V. and Vanosky, J. and van Remortel, N. and Vardaro, M. and Vargas, A. F. and Varma, V. and Vas\'uth, M. and Vecchio, A. and Vedovato, G. and Veitch, J. and Veitch, P. J. and Venneberg, J. and Venugopalan, G. and Verkindt, D. and Verma, P. and Verma, Y. and Veske, D. and Vetrano, F. and Vicer\'e, A. and Vidyant, S. and Viets, A. D. and Vijaykumar, A. and Villa-Ortega, V. and Vinet, J.-Y. and Virtuoso, A. and Vitale, S. and Vo, T. and Vocca, H. and von Reis, E. R. G. and von Wrangel, J. S. A. and Vorvick, C. and Vyatchanin, S. P. and Wade, L. E. and Wade, M. and Wagner, K. J. and Walet, R. C. and Walker, M. and Wallace, G. S. and Wallace, L. and Walsh, S. and Wang, J. Z. and Wang, W. H. and Ward, R. L. and Warner, J. and Was, M. and Washington, N. Y. and Watchi, J. and Weaver, B. and Webster, S. A. and Weinert, M. and Weinstein, A. J. and Weiss, R. and Weller, C. M. and Weller, R. and Wellmann, F. and Wen, L. and We\ss{}els, P. and Wette, K. and Whelan, J. T. and White, D. D. and Whiting, B. F. and Whittle, C. and Wilken, D. and Williams, D. and Williams, M. J. and Williamson, A. R. and Willis, J. L. and Willke, B. and Wilson, D. J. and Winkler, W. and Wipf, C. C. and Wlodarczyk, T. and Woan, G. and Woehler, J. and Wofford, J. K. and Wong, I. C. F. and Wu, D. S. and Wysocki, D. M. and Xiao, L. and Yamamoto, H. and Yang, F. W. and Yang, L. and Yang, Yang and Yang, Z. and Yap, M. J. and Yeeles, D. W. and Yelikar, A. B. and Ying, M. and Yoo, J. and Yu, Hang and Yu, Haocun and Zadro\ifmmode \dot{z}\else \.{z}\fi{}ny, A. and Zanolin, M. and Zelenova, T. and Zendri, J.-P. and Zevin, M. and Zhang, J. and Zhang, L. and Zhang, T. and Zhang, Y. and Zhao, C. and Zhao, G. and Zhao, Yue and Zhou, R. and Zhou, Z. and Zhu, X. J. and Zimmerman, A. B. and Zlochower, Y. and Zucker, M. E. and Zweizig, J.},
  collaboration = {The LIGO Scientific Collaboration and the Virgo Collaboration},
  journal = {Phys. Rev. D},
  volume = {109},
  issue = {2},
  pages = {022001},
  numpages = {45},
  year = {2024},
  month = {Jan},
  publisher = {American Physical Society},
  doi = {10.1103/PhysRevD.109.022001},
  url = {https://link.aps.org/doi/10.1103/PhysRevD.109.022001}
}

@article{Novak:02,
author = {Stephanie Novak and Adolph Moesle},
journal = {J. Lightwave Technol.},
number = {6},
pages = {975},
publisher = {Optica Publishing Group},
title = "{Analytic Model for Gain Modulation in EDFAs}",
volume = {20},
month = {Jun},
year = {2002},
url = {https://opg.optica.org/jlt/abstract.cfm?URI=jlt-20-6-975},
}

@article{Vahlbruchnpro,
    author = {Vahlbruch, H. and Meylahn, F. and Willke, B.},
    title = {Characterization of non-planar ring oscillators at a wavelength of 1064 nm for high precision metrology},
    journal = {Review of Scientific Instruments},
    volume = {96},
    number = {11},
    pages = {113001},
    year = {2025},
    month = {11},
    issn = {0034-6748},
    doi = {10.1063/5.0298918},
    url = {https://doi.org/10.1063/5.0298918},
    eprint = {https://pubs.aip.org/aip/rsi/article-pdf/doi/10.1063/5.0298918/20791148/113001_1_5.0298918.pdf},
}

@article{beam_quality_jitter_AEI,
    author = {Kwee, Patrick and Seifert, Frank and Willke, Benno and Danzmann, Karsten},
    title = {Laser beam quality and pointing measurement with an optical resonator},
    journal = {Review of Scientific Instruments},
    volume = {78},
    number = {7},
    pages = {073103},
    year = {2007},
    month = {07},
    issn = {0034-6748},
    doi = {10.1063/1.2754400},
    url = {https://doi.org/10.1063/1.2754400},
    eprint = {https://pubs.aip.org/aip/rsi/article-pdf/doi/10.1063/1.2754400/15899698/073103_1_online.pdf},
}

@article{CBC_Artemis,
author = {Li-Wei Wei and Fr\'{e}d\'{e}ric Cleva and Catherine Nary Man},
journal = {Opt. Lett.},
number = {24},
pages = {5817--5820},
publisher = {Optica Publishing Group},
title = "{Coherently combined master oscillator fiber power amplifiers for Advanced Virgo}",
volume = {41},
month = {Dec},
year = {2016},
url = {https://opg.optica.org/ol/abstract.cfm?URI=ol-41-24-5817},
doi = {10.1364/OL.41.005817},
}

@article{CBC400,
author = {Felix Wellmann and Nina Bode and Peter Wessels and Ludger Overmeyer and J\"{o}rg Neumann and Benno Willke and Dietmar Kracht},
journal = {Opt. Express},
number = {7},
pages = {10140--10149},
publisher = {Optica Publishing Group},
title = "{Low noise 400 W coherently combined single frequency laser beam for next generation gravitational wave detectors}",
volume = {29},
month = {Mar},
year = {2021},
url = {https://opg.optica.org/oe/abstract.cfm?URI=oe-29-7-10140},
doi = {10.1364/OE.420350},
}

@article{Bode2023,
author = {Nina Bode and Zachary Holmes and Sebastian Ng and Benjamin von Behren and David Ottaway and Benno Willke},
journal = {Opt. Lett.},
number = {17},
pages = {4717--4720},
publisher = {Optica Publishing Group},
title = "{Multiple beam coherent combination via an optical ring resonator}",
volume = {48},
month = {Sep},
year = {2023},
url = {https://opg.optica.org/ol/abstract.cfm?URI=ol-48-17-4717},
doi = {10.1364/OL.500684},
}

@article{multimess,
author = {Péter Mészáros and Derek B. Fox and Chad Hanna and Kohta Murase},
journal = {Nature Reviews Physics},
volume={1},
year={2019},
title="{Multi-messenger astrophysics}",
pages={585--599},
}

@article{Neovan,
author = {Fabian Thies and Nina Bode and Patrick Oppermann and Maik Frede and Bastian Schulz and Benno Willke},
journal = {Opt. Lett.},
number = {3},
pages = {719--722},
publisher = {Optica Publishing Group},
title = "{Nd:YVO4 high-power master oscillator power amplifier laser system for second-generation gravitational wave detectors}",
volume = {44},
month = {Feb},
year = {2019},
url = {https://opg.optica.org/ol/abstract.cfm?URI=ol-44-3-719},
doi = {10.1364/OL.44.000719},
}

@Article{LIGO_O3_laser,
AUTHOR = {Bode, Nina and Briggs, Joseph and Chen, Xu and Frede, Maik and Fritschel, Peter and Fyffe, Michael and Gustafson, Eric and Heintze, Matthew and King, Peter and Liu, Jian and Oberling, Jason and Savage, Richard L. and Spencer, Andrew and Willke, Benno},
TITLE = "{Advanced LIGO Laser Systems for O3 and Future Observation Runs}",
JOURNAL = {Galaxies},
VOLUME = {8},
YEAR = {2020},
NUMBER = {4},
ARTICLE-NUMBER = {84},
pages = {84},
URL = {https://www.mdpi.com/2075-4434/8/4/84},
ISSN = {2075-4434},
DOI = {10.3390/galaxies8040084}
}

@article{LIGO_O4_laser,
author = {Nina Bode and Fabian Meylahn and Benno Willke},
journal = {Opt. Express},
number = {20},
pages = {29469--29478},
publisher = {Optica Publishing Group},
title = "{Sequential high power laser amplifiers for gravitational wave detection}",
volume = {28},
month = {Sep},
year = {2020},
url = {https://opg.optica.org/oe/abstract.cfm?URI=oe-28-20-29469},
doi = {10.1364/OE.401826},
}

@article{Kwee2012,
author = {P. Kwee and C. Bogan and K. Danzmann and M. Frede and H. Kim and P. King and J. P\"{o}ld and O. Puncken and R. L. Savage and F. Seifert and P. Wessels and L. Winkelmann and B. Willke},
journal = {Opt. Express},
number = {10},
pages = {10617--10634},
publisher = {Optica Publishing Group},
title = "{Stabilized high-power laser system for the gravitational wave detector advanced LIGO}",
volume = {20},
month = {May},
year = {2012},
url = {https://opg.optica.org/oe/abstract.cfm?URI=oe-20-10-10617},
doi = {10.1364/OE.20.010617},
}

@ARTICLE{triangular_cavity,
	author = {Skettrup, Torben and Meelby, Thomas and Færch, Kjartan and Frederiksen, Søren Lund and Pedersen, Christian},
	title = {Triangular laser resonators with astigmatic compensation},
	year = {2000},
	journal = {Applied Optics},
	volume = {39},
	number = {24},
	pages = {4306 – 4312},
	doi = {10.1364/AO.39.004306},
	url = {https://www.scopus.com/inward/record.uri?eid=2-s2.0-0001747794&doi=10.1364%2fAO.39.004306&partnerID=40&md5=cd0bac42a11be0548381cb3a7b2be201},
	type = {Article},
	publication_stage = {Final},
	source = {Scopus},
	note = {Cited by: 16}
}

@article{Wellmann2019,
author = {Felix Wellmann and Michael Steinke and Fabian Meylahn and Nina Bode and Benno Willke and Ludger Overmeyer and J\"{o}rg Neumann and Dietmar Kracht},
journal = {Opt. Express},
number = {20},
pages = {28523--28533},
publisher = {Optica Publishing Group},
title = {High power, single-frequency, monolithic fiber amplifier for the next generation of gravitational wave detectors},
volume = {27},
month = {Sep},
year = {2019},
url = {https://opg.optica.org/oe/abstract.cfm?URI=oe-27-20-28523},
doi = {10.1364/OE.27.028523},
}

@article{Wellmann2020,
author = {Felix Wellmann and Michael Steinke and Peter Wessels and Nina Bode and Fabian Meylahn and Benno Willke and Ludger Overmeyer and J\"{o}rg Neumann and Dietmar Kracht},
journal = {Appl. Opt.},
number = {26},
pages = {7945--7950},
publisher = {OSA},
title = {Performance study of a high-power single-frequency fiber amplifier architecture for gravitational wave detectors},
volume = {59},
month = {Sep},
year = {2020},
url = {http://opg.optica.org/ao/abstract.cfm?URI=ao-59-26-7945},
doi = {10.1364/AO.401048},
}

@article{Dixneuf2020,
author = {Cl\'{e}ment Dixneuf and Germain Guiraud and Yves-Vincent Bardin and Quentin Rosa and Mathieu Goeppner and Ad\`{e}le Hilico and Christophe Pierre and Johan Boullet and Nicholas Traynor and Giorgio Santarelli},
journal = {Opt. Express},
number = {8},
pages = {10960--10969},
publisher = {OSA},
title = "{Ultra-low intensity noise, all fiber 365 W linearly polarized single frequency laser at 1064 nm}",
volume = {28},
month = {Apr},
year = {2020},
url = {http://opg.optica.org/oe/abstract.cfm?URI=oe-28-8-10960},
doi = {10.1364/OE.385095},
}

@INPROCEEDINGS{ClevaPstab,
author={Cleva, Frédéric and Coulon, Jean-Pierre and Wei, Li Wei and Turconi, Margherita and Merzougui, Mourad and Genin, Éric and Pillant, Gabriel and Kéfélian, Fabien},
booktitle={2021 Conference on Lasers and Electro-Optics Europe and European Quantum Electronics Conference (CLEO/Europe-EQEC)},
title="{Laser power stabilization for Advanced VIRGO}",
year={2021},
volume={},
number={},
pages={1-1},
doi={10.1109/CLEO/Europe-EQEC52157.2021.9542330}}

@article{ISYS,
doi = {10.1088/0264-9381/19/7/381},
url = {https://doi.org/10.1088/0264-9381/19/7/381},
year = {2002},
month = {mar},
publisher = {},
volume = {19},
number = {7},
pages = {1829},
author = {F Bondu and A Brillet and F Cleva and H Heitmann and M Loupias and C N Man and H Trinquet and the VIRGO Collaboration},
title = "{The VIRGO injection system}",
journal = {Classical and Quantum Gravity}
}

@article{SBS,
    author = "{M. Hildebrandt et al.}",
    title = "{Brillouin Scattering Spectra in High-Power Single-Frequency Ytterbium Doped Fiber Amplifiers}",
    journal = "Opt. Express",
    volume = "16",
    pages = "15970--15979",
    year = "2008",
}

@ARTICLE{SBS_noise,
  author={Peral, E. and Yariv, A.},
  journal={IEEE Journal of Quantum Electronics}, 
  title="{Degradation of modulation and noise characteristics of semiconductor lasers after propagation in optical fiber due to a phase shift induced by stimulated Brillouin scattering}", 
  year={1999},
  volume={35},
  number={8},
  pages={1185-1195},
  doi={10.1109/3.777219}}

@article{Drever1983,
  author  = {R. W. P. Drever and J. L. Hall and F. V. Kowalski and J. Hough and G. M. Ford and A. J. Munley and H. Ward},
  title   = {Laser phase and frequency stabilization using an optical resonator},
  journal = {Applied Physics B},
  year    = {1983},
  volume  = {31},
  number  = {2},
  pages   = {97--105},
  doi     = {10.1007/BF00702605},
  url     = {https://doi.org/10.1007/BF00702605},
  issn    = {1432-0649}
}

@techreport{Canuel2013,
  author       = {B. Canuel and E. Genin and J. Marque and P. Benoit},
  title        = {Reference Cavity characterization},
  institution  = {Virgo Collaboration},
  year         = {2013},
  number       = {VIR-0231A-13},
  type         = {Scientific and Technical Note},
  url          = {https://tds.virgo-gw.eu/ql/?c=9577},
  note         = {Issued May 20, 2013}
}

@techreport{ET2020,
  title        = "{ET design report update 2020}",
  institution  = {ET Steering Committee},
  year         = {2020},
  type         = {Technical Report},
  url          = {https://www.einsteintelescope-emr.eu/wp-content/uploads/2024/05/ET-0007B-20_ETDesignReportUpdate2020.pdf&ved=2ahUKEwiS6eKv6pWUAxWNRaQEHYmcNigQFnoECAwQAQ&usg=AOvVaw0HaehoGwV4svC9qe15DyDT}
}

@phdthesis{Day1991,
  author  = {T. Day},
  title   = {Frequency-stabilized solid state lasers for coherent optical communications},
  school  = {Stanford University},
  year    = {1991},
  address = {Stanford, CA},
  type    = {Ph.D. thesis}
}

@article{DCreadout,
  title = "{DC-readout of a signal-recycled gravitational wave detector}",
  author={Hild, Stefan and Grote, Hartmut and Degallaix, Jerome and Chelkowski, Szczepan and Danzmann, Karsten and Freise, Andreas and Hewitson, Martin and Hough, J and L{\"u}ck, Harald and Prijatelj, M and others},
  journal={Classical and Quantum Gravity},
  volume={26},
  number={5},
  pages={055012},
  year={2009}
}

@article{Kwee2007,
    author = {Kwee, Patrick and Seifert, Frank and Willke, Benno and Danzmann, Karsten},
    title = {Laser beam quality and pointing measurement with an optical resonator},
    journal = {Review of Scientific Instruments},
    volume = {78},
    number = {7},
    pages = {073103},
    year = {2007},
    month = {07},
    issn = {0034-6748},
    doi = {10.1063/1.2754400},
    url = {https://doi.org/10.1063/1.2754400},
    eprint = {https://pubs.aip.org/aip/rsi/article-pdf/doi/10.1063/1.2754400/15899698/073103_1_online.pdf},
}

@article{Millo2014,
author = {Jacques Millo and Mourad Merzougui and Sibilla Di Pace and Walid Chaibi},
journal = {Appl. Opt.},
number = {32},
pages = {7761--7772},
publisher = {Optica Publishing Group},
title = {High bandwidth frequency lock of a rigid tunable optical cavity},
volume = {53},
month = {Nov},
year = {2014},
url = {https://opg.optica.org/ao/abstract.cfm?URI=ao-53-32-7761},
doi = {10.1364/AO.53.007761},
}

@article{Kwee2009,
author = {Patrick Kwee and Benno Willke and Karsten Danzmann},
journal = {Opt. Lett.},
number = {19},
pages = {2912--2914},
publisher = {Optica Publishing Group},
title = {Shot-noise-limited laser power stabilization with a high-power photodiode array},
volume = {34},
month = {Oct},
year = {2009},
url = {https://opg.optica.org/ol/abstract.cfm?URI=ol-34-19-2912},
doi = {10.1364/OL.34.002912},
}

\end{document}